%% file: Amaya-Almazan_et_al.tex
\documentclass[twocolumn,trackchanges]{aastex62}

\usepackage{nccmath}
\usepackage{multirow}
\usepackage{amsmath, amstext}
\usepackage{amssymb}
\usepackage{amsmath}
\usepackage{mathtools}
\usepackage{color} 
\usepackage{threeparttable}
\usepackage{float}
\usepackage{placeins}
\usepackage[utf8]{inputenc}

\def\mathbi#1{\textbf{\em #1}}

\received{...}
\revised{...}
\accepted{...}

\submitjournal{ApJ}

\shorttitle{Multiwavelength analysis of B2~1633+382}
\shortauthors{Amaya-Almaz\'an et al.}

\begin{document}

\title{Multiwavelength analysis and the C~IV $\lambda1549$~\AA\ emission line behavior from 2008 to 2020 of the FSRQ B2~1633+382}

\correspondingauthor{Ra\'ul A. Amaya-Almaz\'an}
\email{rantonioaa@inaoep.mx, amayaalmazanra@gmail.com}

\author[0000-0002-9443-7523]{Ra\'ul A. Amaya-Almaz\'an}
\affil{Instituto Nacional de Astrof\'isica, \'Optica y Electr\'onica, Luis Enrique Erro $\# 1$, 
Tonantzintla, Puebla 72840, M\'exico}

\author[0000-0002-2558-0967]{Vahram Chavushyan}
\affil{Instituto Nacional de Astrof\'isica, \'Optica y Electr\'onica, Luis Enrique Erro $\# 1$,
Tonantzintla, Puebla 72840, M\'exico}

\author[0000-0002-5442-818X]{Victor M. Pati\~no-\'Alvarez}
\affil{Instituto Nacional de Astrof\'isica, \'Optica y Electr\'onica, Luis Enrique Erro $\# 1$, 
Tonantzintla, Puebla 72840, M\'exico}
\affiliation{Max-Planck-Institut f\"ur Radioastronomie, Auf dem H\"ugel 69,
D-53121 Bonn, Germany}

\begin{abstract}

The flat-spectrum radio quasar B2~1633+382 (4C~38.41) has been monitored for several years and has presented correlated variability in multiple wavelengths. In this article, we are performing different analyses for multiple frequencies, from gamma-rays to radio, as well as, the C~IV $\lambda$1549~\AA\ emission line and the $\lambda$1350~\AA\ continuum. Using the non-thermal dominance parameter, we separated the C~IV and the continuum light curves for when the dominant source of continuum is the accretion disk or the jet. We found a correlation at a delay consistent with zero between the line and the continuum dominated by disk emission indicating a very small broad-line region (BLR). From the resulting delay between the 15~GHz and gamma-rays, we estimated the distance of the gamma-ray emission region from the jet apex to be $\sim37$ pc. The C~IV flux decreases when the continuum and gamma-rays increase at some of the high activity periods. The C~IV profile presents a larger variable component in its blue wing. The relation between the luminosities of C~IV and the continuum does not completely follow the relation for a quasar sample. Our results lead us to propose an outflow of BLR material in the jet-flow direction, a gamma-ray production through magnetic reconnection for the flaring event of mid-2011, and that there is not enough BLR material close to the radio core to be easily ionized by the non-thermal continuum.

\end{abstract}


\section{Introduction} \label{sec:intro}

Blazars are a type of radio-loud active galactic nuclei (AGN) which have their relativistic jet pointed towards Earth \citep{Urry&Padovani1995} triggering Doppler boosting \citep{Sher1968}. They present variability across multiple wavebands and different time periods \citep[][and references therein]{Fan2018,Gupta2018}. The variability of flat spectrum radio quasars (FSRQs) can be as fast as within a day (intraday variability, IDV) to slower variations of months or years. This type of blazar is characterized by showing prominent emission lines, emulating the spectra of quasars. Their spectral energy distribution (SED) displays a low- and a high-energy components. The low-energy emission goes from radio to X-rays with its origin being thermal, from the accretion disk, and non-thermal, from the jet through synchrotron emission. Meanwhile, the high-energy emission ranges from X-rays to gamma-rays with their production having place in the jet through inverse Compton (IC) scattering of low-energy photons \citep[e.g.,][]{Bottcher2007,Bottcher2013,Romero2017}. These seed photons may come to the jet from different sources, if they come from the jet itself the process is called synchrotron self-Compton \citep[SSC,][]{Bloom1996}. However, if they come from external sources, such as the accretion disk, the broad-line region (BLR), or the dusty torus, the process is named external inverse Compton \citep[EC,][]{Sikora1994}.

The object of interest of this study, B2~1633+382 (4C~38.41), has been constantly monitored through observational campaigns for several years. This object is a typical FSRQ where multiple frequencies and optical polarization present correlated variability \citep{Raiteri2012,Algaba2018_1,HagenThorn2019,Wang2021,Pandey2022}. It presents structural and flux variability in its parsec scale jet \citep[e.g.,][]{Jorstad2017,Algaba2018_2}. \cite{Liu2010} found helical motions of jet components and that their ejection angle is time dependent, they interpreted these results in the context of binary black hole (BBH) models. Supporting this, periodicities have been found in the radio \citep{Fan2007}, in the optical and hints in the gamma-rays \citep{OteroSantos2020}. However, \cite{OteroSantos2020} discussed that these periodicities may be explained by geometrical effects such as helical structures or shock fronts \citep{MarscherGear1985}. Furthermore, \cite{Algaba2019} found that the jet of this source shows a significant bend within the first milliarcseconds. They discussed that this bend could be the result from the interaction of the jet with the ambient medium (cold gas) or the projection in the sky of a helical or harmonic jet trajectory, due to the presence of helical magnetic fields and/or a BBH.

The location of the gamma-ray emission region in this type of sources is still under debate. The first possible location is close to the supermassive black hole (SMBH) within the central parsec \citep[e.g.,][]{FinkeDermer2010,Tavecchio2010,PoutanenStern2010}. The second possible location is further downstream the jet \citep[e.g.,][]{Sikora2009,LeonTavares2011,PatinoAlvarez2019}. However, there are sources where there might be multiple gamma-ray emission regions \citep[e.g.][]{PatinoAlvarez2018,AmayaAlmazan2021}. For B2~1633+382, \cite{Algaba2018_1} speculated that the gamma-ray emission region responsible for large gamma-ray flares is located at $\sim 1$ pc from the jet apex. Meanwhile, \cite{Wang2021} found it to be located at $\sim 27$ pc from the base of the jet. \cite{Algaba2018_1} also suggest that the less powerful and faster flares may be originated in a smaller region within the central parsec.
However, \cite{Raiteri2012} proposed that a change in the Doppler factor induced by variations in the viewing angle at parsec scales could explain the flaring behavior.

The relation between the luminosities of the UV-continuum and the Mg~II emission line has been studied recently. \cite{Chavushyan2020} and \cite{AmayaAlmazan2021} compared the intrinsic relation for CTA~102 and 3C~454.3, respectively, to the \cite{Shen2011} relation estimated for a quasar sample dominated by radio-quiet sources. They showed that the intrinsic relation for these strong radio-loud sources does not completely follow the relation of \cite{Shen2011} due to the UV-continuum luminosity being contaminated by jet emission. They discussed that this contamination of the UV-continuum would also affect the black hole mass estimation for blazar-type sources from single-epoch and reverberation mapping techniques.

In order to explain the relationship between the optical continuum, the emission-line variability, and the kinematics of the jet in sub-parsec scales, the existence of BLR material close to the jet was proposed \citep{Arshakian2010,LeonTavares2010}. The non-thermal continuum from the jet ionizes this BLR material causing the emission-line flare-like behaviors. The first observational evidence of the relation between the emission-line variability and the non-thermal continuum was presented by \cite{Leon-Tavares2013} and later confirmed \citep{Isler2013,Jorstad2013} for the FSRQ 3C 454.3. Recently, the same type of interaction was reported by \cite{Chavushyan2020} in CTA 102 but in a more violent event. There are different models trying to explain the origin of these BLR clouds far from the accretion disk. One of the models proposes that these clouds arise from winds of the accretion disk accelerated by the magnetic fields of the jet \citep{Perez1989,PaltaniTurler2003,FinkeDermer2010}. Another possible origin could be the interaction between the jet and a red giant star; the jet pulling out the outer layers of the star, and in this way enriching the medium with material to be ionized \citep{Bosch_Ramon2012,Khangulyan2013}. A similiar model proposes that the interaction between the jet and a star-forming region is the agent providing the line-emitting material \citep{Zacharias2019}.

In this work, we are analyzing the variability of multiple wavebands along with the $\lambda1350$~\AA\ continuum and the C~IV $\lambda$1549~\AA\ emission-line extracted from spectra taken at the Steward Observatory \citep{Smith2009}. This is the first study combining these features for a long observation period (2008-2020). Moreover, we are analyzing the relationship between the UV-continuum and emission-line luminosities. These analyses were also performed separating the UV-continuum by its dominant source (accretion disk or jet) with the aid of the non-thermal dominance (NTD) parameter \citep{Shaw2012,PatinoAlvarez2016}. Finally, we are discussing the implications of our results regarding the gamma-ray emission-region location and origin.

The cosmological parameters adopted throughout this paper are H$_0=71$ km s$^{-1}$ Mpc$^{-1}$, $\Omega_{\Lambda}=0.73$, $\Omega_m=0.27$. At the redshift of the source, z $=1.814$ \citep{Paris2017}, the spatial scale of 1\arcsec\ corresponds to a physical scale of 8.5 kpc and the luminosity distance is 13.943 Gpc.

\section{Observations}\label{sec:observe}

\subsection{Optical spectra}\label{ssec:op-spec}

We used 364 optical spectra from the data collection of the Ground-based Observational Support of the Fermi Gamma-Ray Space Telescope at the University of Arizona from mid-2008 through late 2018. These spectra are calibrated against the V-band magnitude \citep{Smith2009}. We took the spectra to the rest frame of the object and applied a cosmological correction of the form $(1+z)^3$ to the flux. Since we are only interested in the relative changes of flux, galactic reddening correction was not applied.
The flux of the C~IV $\lambda$1549~\AA\ emission line of each spectrum was measured, for which we needed to subtract the He II $\lambda$1640, O III] $\lambda$1663 emissions, as well as, the continuum. It is worth noting that the spectra did not show a significant Fe II emission and because of this we did not take this emission into account for the procedure. We performed the fitting procedure with the aid of the framework \texttt{astropy.modeling}\footnote{\url{https://docs.astropy.org/en/stable/modeling/index.html}}.

\begin{figure}[htbp]
\begin{center}
\includegraphics[width=0.48\textwidth]{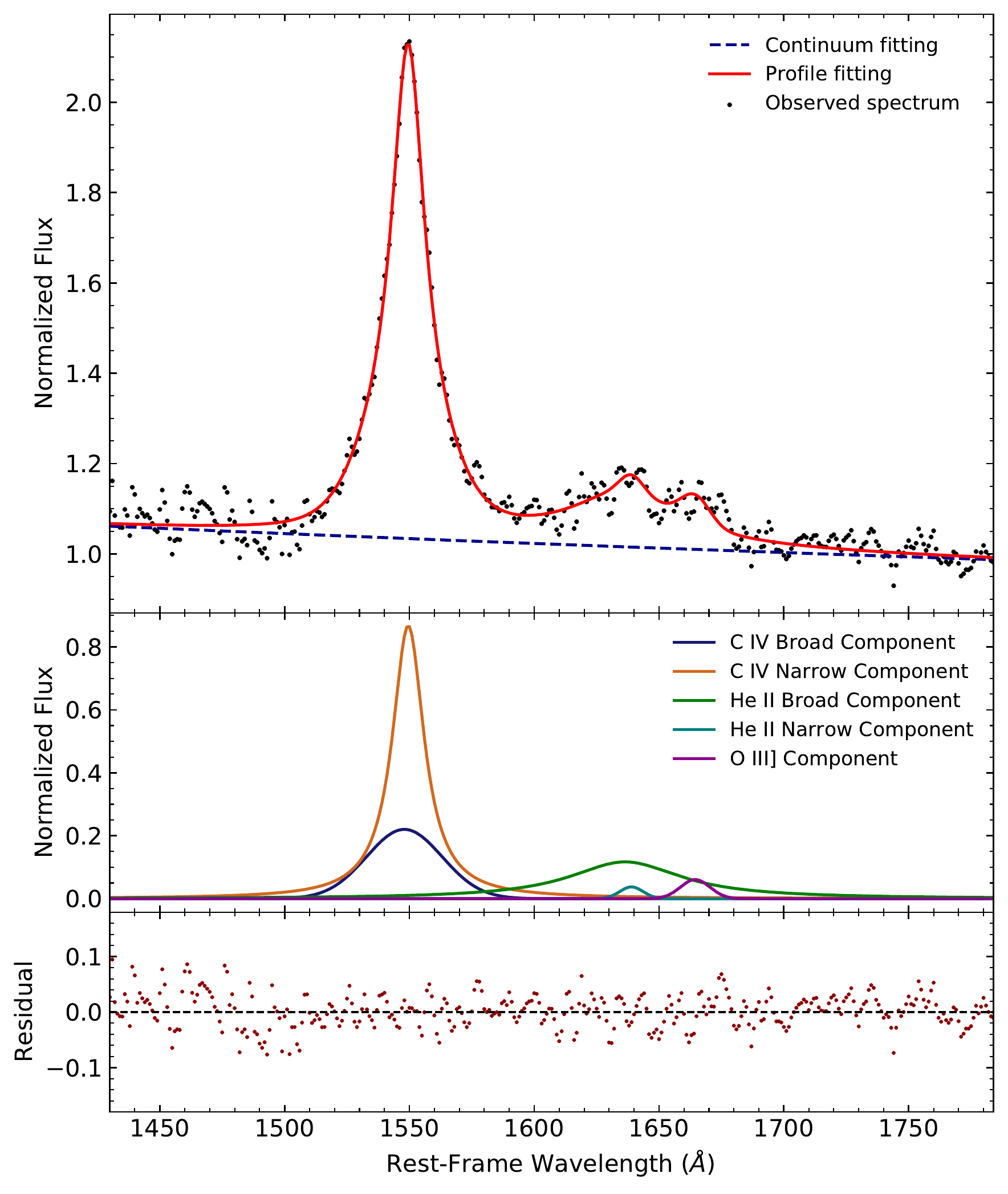}
\caption{Example decomposition of the C~IV $\lambda$1549~\AA\ emission line in a low continuum state spectrum taken at the Steward Observatory on May 18th, 2012. Top panel: Observed spectrum with the obtained best fit, a power-law function as the continuum is depicted. Middle panel: The broad and narrow components used to fit the C~IV $\lambda$1549~\AA\, as well as, the He II and O III] components. Bottom panel: The residuals from the subtraction of the best fit to the observed spectrum.}
\label{spec-decomp}
\end{center}
\end{figure}

First, we fitted the continuum with a power-law function and subtracted it from the spectrum. Then, we fitted the C~IV and He~II lines with a narrow plus broad component each and the O III] with a single component. The narrow and broad components refers to their relative width rather than the emission-line regions. We subtracted the continuum and the modeled profile of He II and O III] emission lines to obtain a clean spectrum. Finally, we integrated the clean spectrum in the range of $1500-1600$~\AA\ to estimate the C~IV emission line flux. We also estimated the flux of the $\lambda$1450~\AA\ continuum by taking the mean value in the range $1430-1480$~\AA\ of the spectrum without the C~IV contribution. The Figure~\ref{spec-decomp} shows an example of the performed spectral decomposition. The uncertainty in the C~IV flux measurement was measured following the procedure of \cite{Tresse1999}, which is produced by the dispersion and the signal-to-noise ratio of the spectra. For the estimation of the $\lambda$1450~\AA\ continuum flux uncertainty we calculated the standard deviation in the range specified above.

\input{Table1_flux}

\begin{figure}[htbp]
\begin{center}
\includegraphics[width=0.48\textwidth]{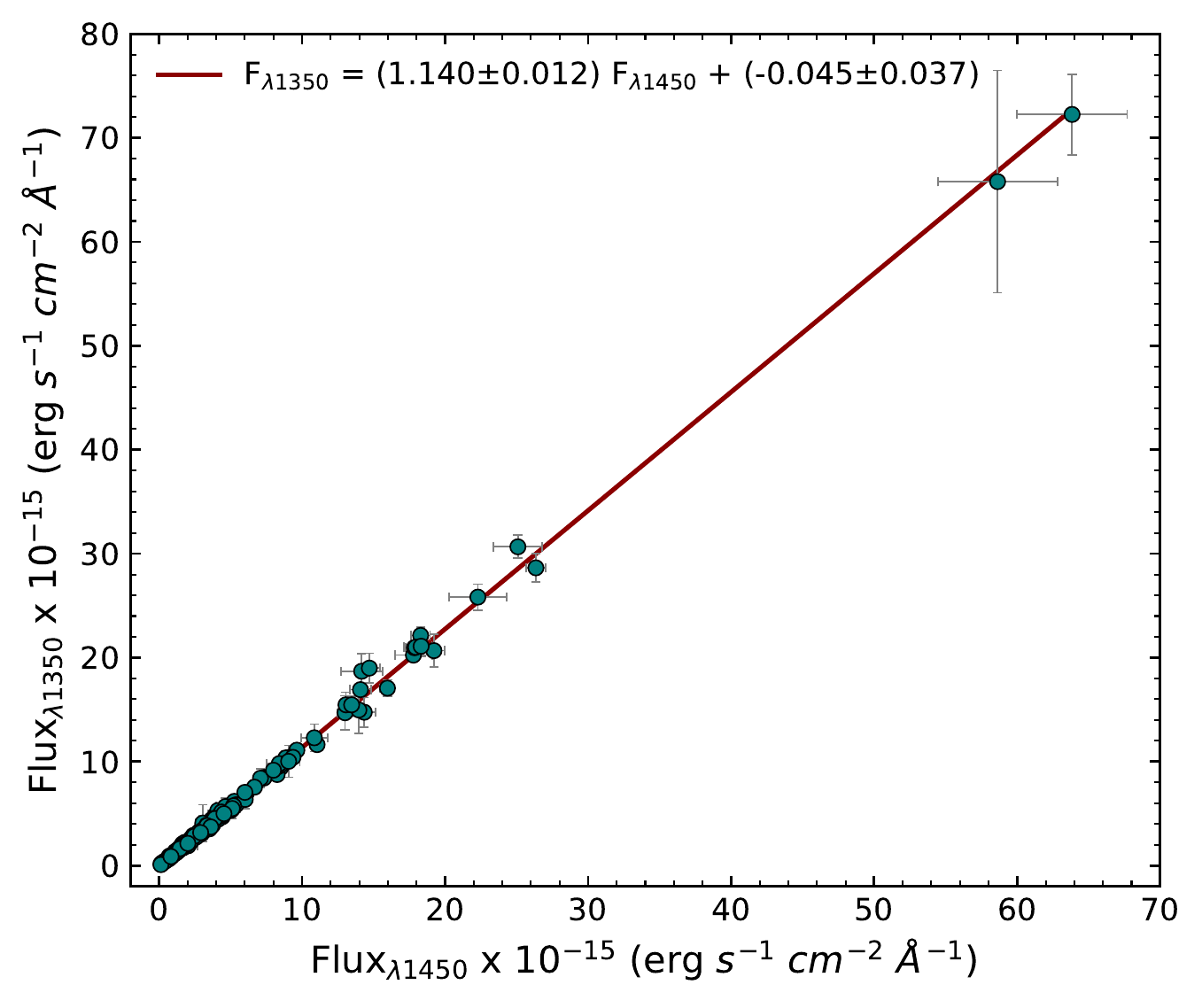}
\caption{Relation between the $\lambda$1450~\AA\ and $\lambda$1350~\AA\ continuum fluxes for the Roma-BZCAT blazar sample.}
\label{1450vs1350}
\end{center}
\end{figure}

Our luminosity analysis involves the relation between the $\lambda$1350~\AA\ continuum and the C~IV emission line estimated by \cite{Shen2011} for a non-blazar sample. However, this continuum wavelength was not available in the observed optical spectra. Since the $\lambda$1450~\AA\ flux is available, we proceeded to look for a relation between the $\lambda$1350~\AA\ and the $\lambda$1450~\AA\ so that we could estimate the former from the latter. We used sources from the Roma-BZCAT catalog \citep{Massaro2009,Massaro2015} of blazars of the FSRQ type that had spectra available in the Sloan Digital Sky Survey DR16 database. The selected objects needed to cover the $\lambda$1350~\AA\ and $\lambda$1450~\AA\ continuum windows. Hence, we set redshift limits ($1.73-5.12$) to meet these conditions. The resultant sample consists of 152 objects. We measured the flux values for $\lambda$1350~\AA\ and $\lambda$1450~\AA\ for each object by taking the mean value around the desired wavelengths ($\pm 50$~\AA) and using the standard deviation as the uncertainty. We performed linear fitting for this relation using Orthogonal Distance Regression from the \texttt{SciPy ODR package}\footnote{\url{https://docs.scipy.org/doc/scipy/reference/odr.html}}. The fit is statistically significant at a significance level of 95\%, considering the p-value ($p_v$, the probability of obtaining a chi-square value equal or higher, by chance) of $<0.05$. The obtained relation between the two wavelengths is shown in Figure~\ref{1450vs1350}. Finally, with the obtained linear relation, we estimated the $\lambda$1350~\AA\ continuum for B2~1633+382 spectra. The fluxes of the $\lambda$1350~\AA\ continuum and the C~IV $\lambda$1549~\AA\ emission line are shown in Table~\ref{Table_flux}.

With the $\lambda$1350~\AA\ continuum and the C~IV emission-line flux, we estimated the non-thermal dominance parameter \citep[NTD,][]{Shaw2012}. We followed the procedure in \cite{PatinoAlvarez2016}. This parameter allows us to determine when the continuum is dominated by the non-thermal emission (the jet as the source). If NTD $=1$, the emission from the continuum is exclusively thermal, with the accretion disk as the source. If the NTD is between 1 and 2, the accretion disk is the main source of the continuum, but the jet also contributes to the emission. For NTD $=2$, the contributions from the disk and the jet are equal. Finally, for values of NTD $>2$, the non-thermal emission from the jet dominates. The light curve of this parameter is shown in panel j) of Figure~\ref{multiwavelength}.

\subsection{Photometric observations}

\begin{figure*}[htbp]
\begin{center}
\includegraphics[width=1\textwidth]{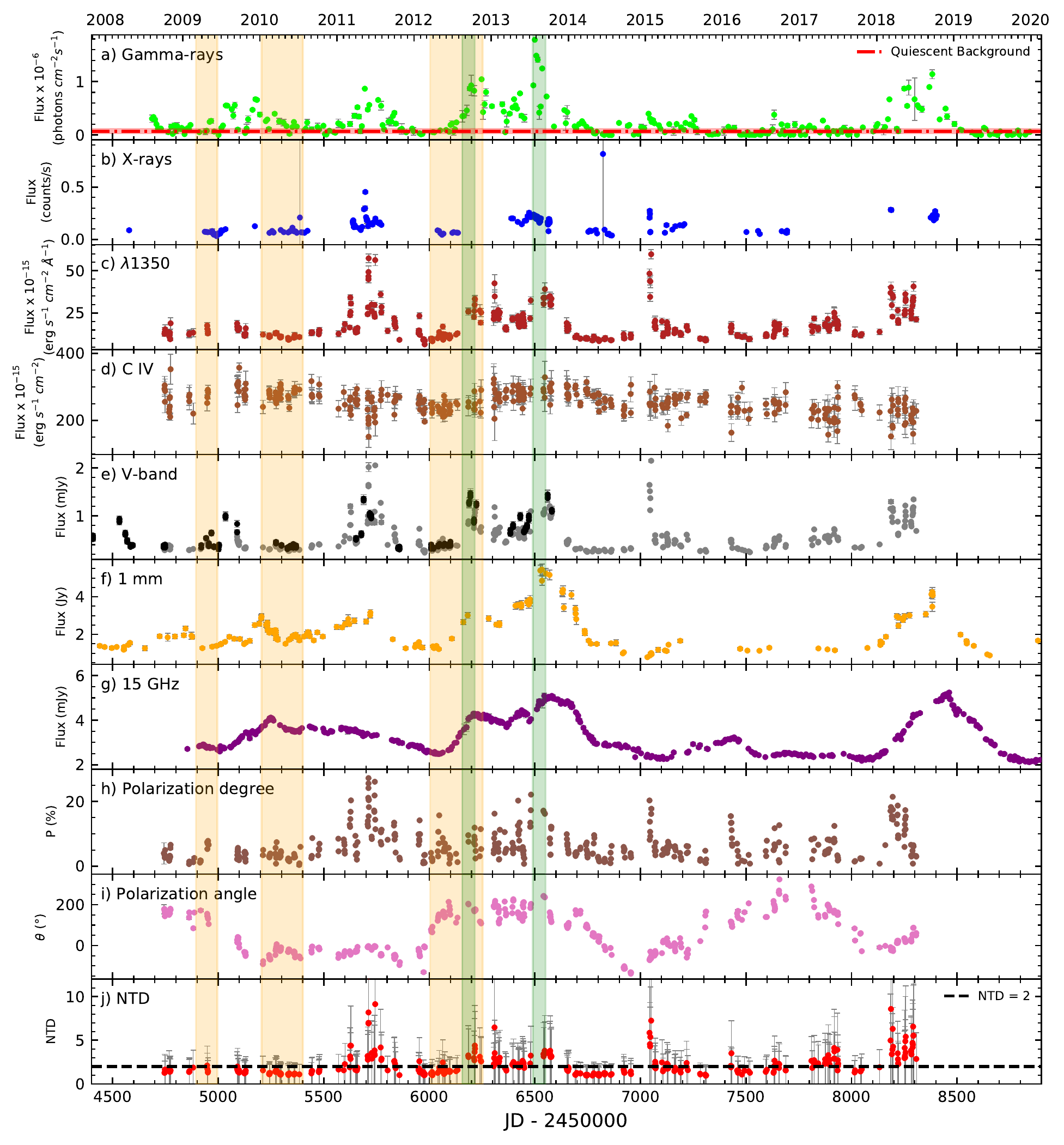}
\caption{Multiwavelength light curves for B2~1633+382 (see Section \ref{sec:observe} for details on the observations). The vertical orange and green stripes represent the ejection of components from the 43 GHz core detected by \cite{Jorstad2017} and \cite{Algaba2018_2}, respectively. The width of these stripes represent their respective uncertainty. Note: The Catalina Surveys V-band data is represented by the black points in panel e).}
\label{multiwavelength}
\end{center}
\end{figure*}

Using the data from the \textit{Fermi} Large Area Telescope \citep[LAT,][]{Abdo2009} database we built the gamma-ray light curve in the energy range $0.1-300$ GeV with the \texttt{Fermitools version 1.0.2}. From the 4FGL catalog \citep{Abdollahi2020}, all sources within 15 degrees of the location of B2~1633+382 were included in the model. The parameters for the sources within 5 degrees were left free; while for the remaining sources, only the normalization was left free.
The Swift X-Ray Telescope (XRT) data were processed using the SWIFT tools (Swift Software version 3.9, \texttt{FTOOLS version 6.12} and \texttt{XSPEC version 12.7.1}) and the light curves were generated with \texttt{xrtgrblc version 1.6}. Details of the reduction procedure can be found in \cite{Stroh2013}.
From the Ground-based Observational Support of the Fermi Gamma-Ray Space Telescope at the University of Arizona \citep[Steward Observatory,][]{Smith2009}, we took the V-band and R-band data, as well as, the optical linear polarization degree and the optical polarization angle data (from 5000 - 7000 \AA, observed frame). In order to account for $180\degr$ wraps in the polarization angle ($\theta$), we considered the data with $\theta/\sigma_{\theta} > 5$ and allowed for changes $\leq 90\degr$ between consecutive observations assuming minimal variability. Additionally, to complement the V-band data we used the Catalina Surveys data available \citep{CatalinaRef}. 
From the Sub-Millimeter Array (SMA) public database, the 1~mm data were retrieved. Details on the observations and data reduction can be found in \cite{Gurwell2007}.
The 15~GHz data came from the public database of the Owens Valley Radio Observatory \citep[OVRO,][]{Richards2011}.
The light curves for the described data are shown in Figure~\ref{multiwavelength}.

\section{Cross-correlation analysis} \label{sec:crosscorr}

The relation between the different bands was determined through the cross-correlation analysis. Following the methodology described in the Appendix \ref{apB}, three different methods were used to ensure the validity of the results. Only time delays with a correlation coefficient at greater than 99\% significance that appeared in at least two of the three methods were taken into account. In every case, the delay reported is the average of the results from the two or three methods.  We also took into account possible spurious correlations, by analyzing the Fourier Transform and power spectrum of the individual light curves. According to \cite{Bath1974}, the cross-correlation function forms a Fourier pair with the power spectrum, which lead us to expect that any periods with high spectral power, obtained from the power spectrum, can reflect on the cross-correlation function, as a high correlation coefficient. Therefore, after obtaining all the delays with a correlation at 99\% significance or higher, we check in the Fourier transform and the power spectra if there is high power in any of these delays (by inverting the frequency, we can have the x-axis in time units). Any such delays are considered spurious correlations, since they are derived from the behavior of an individual light curve and not from the relationship between two light curves. Using the same method, we check for spurious correlations derived from the cadence of observations, by constructing unitary light curves (interpolated light curves with a value of 1, where there is an observation, and 0 everywhere else); by doing this, we ensure that any points of high spectral power, are caused only by the sampling of the light curve, and not by its variability. The delay uncertainty reported is the largest of the three methods. The order in which the cross-correlations were performed corresponds to the order in which the results will be presented. This means that a positive delay corresponds to the first band leading the second and a negative delay corresponds to when the first band is lagging the second.

We found a delay of $-69.5\pm8.7$ days between the 15~GHz and the gamma-rays, showing that the 15~GHz emission region is not co-spatial with the high energy emission region. The gamma-rays correlate, as well, with 1~mm at a delay of $19.3\pm16.3$ days. 

\begin{figure*}[htbp]
\begin{center}
\includegraphics[width=1\textwidth]{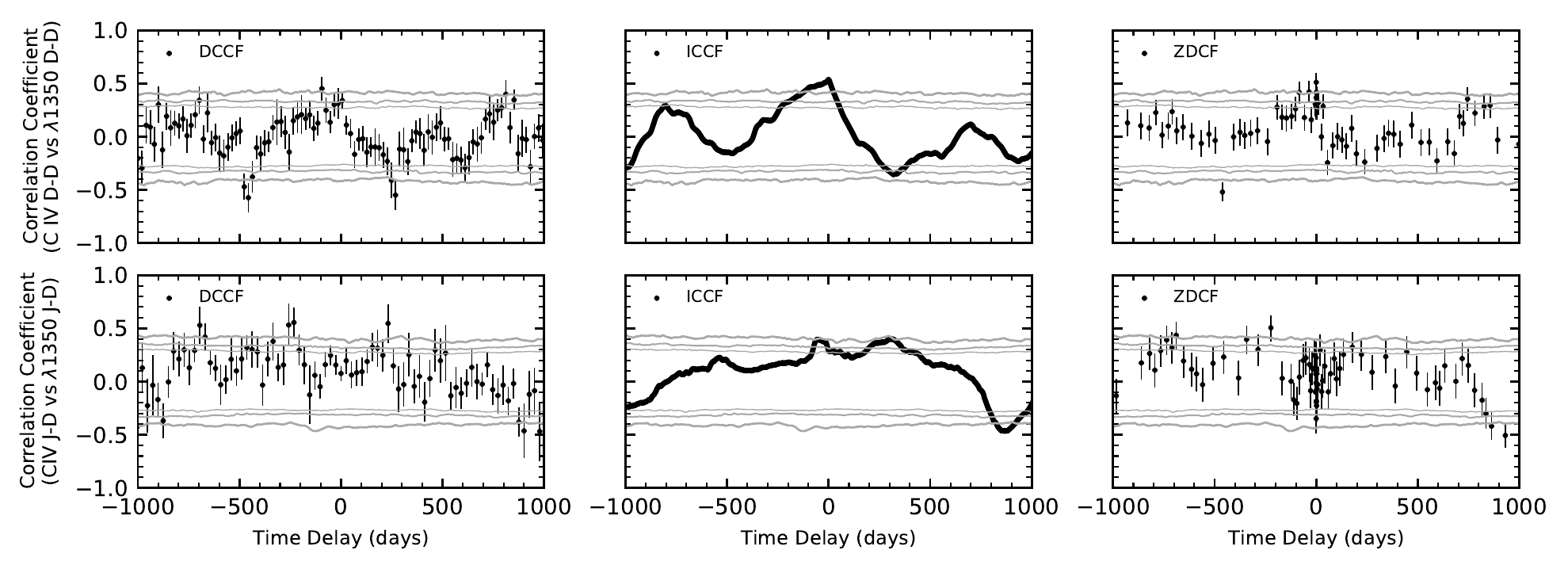}
\caption{Cross-correlation figures for the $\lambda$1350~\AA\ continuum and the C~IV $\lambda$1549~\AA\ emission-line light curves separated by disk and jet domination as described in Section \ref{sec:crosscorr}. The top row is for the disk-dominated light curves and the bottom for the jet-dominated. Each column represents the method used for the analysis. The first method is the Discrete Cross-Correlation Function (DCCF), the second is the Interpolation method (ICCF), and the third method is the Z-transformed Cross-Correlation Function (ZDCF). The significance at the 90\%, 95\%, and 99\% are represented by grey lines.}
\label{cc-ntd}
\end{center}
\end{figure*}

Additionally, there is a delay of $-7.4\pm8.7$ days between the continuum and gamma-rays consistent with zero. This delay of zero is confirmed by the delay of $0.0\pm5.9$ and $0.7\pm8.7$ days of the V-band with the continuum and the gamma-rays, respectively. This indicates co-spaciality between the regions that emit the seed photons needed for the gamma-ray production, the continuum, and the V-band.

We could not find a delay with a significant correlation for the C~IV against the $\lambda$1350~\AA\ continuum, V-band or gamma-rays. To further investigate the emission-line behavior, we separated the continuum and C~IV light curves in disk-dominated (D-D, $1<$~NTD~$<2$) and jet-dominated (J-D, NTD~$>2$). Then, we performed the cross-correlation analysis for these D-D and J-D light curves. 

We found that there is a delay of $0.0\pm10.1$ days between C~IV and the continuum for the D-D light curves. For the J-D light curves, we found a delay of $-227.4\pm12.9$ days. These results are shown in Figure~\ref{cc-ntd}. We tested whether these correlations were spurious due to sampling effects by inspecting the power spectrum of the unitary light curves with the same cadence as the D-D and J-D light curves. The test results showed that the correlations are real.

The rest of the results of the cross-correlation analysis for the full light curves are listed in Table~\ref{Table_CC}. All cross-correlation figures are displayed in the Appendix \ref{appendix}.

\input{Table2_cc}

\section{Variability}\label{sec:var}

\subsection{Photometric variability}

The light curves of the different wavelengths show strong variability during the studied period (2008-2020). There are several major flaring events in gamma-rays and multiple ejections from the 43 GHz radio core. These ejections reported by \cite{Jorstad2017} and \cite{Algaba2018_2} are shown in Figure~\ref{multiwavelength} in orange and green vertical stripes, respectively.

The gamma-ray emission can be separated in quiescent level flux and the flaring state flux. We estimated the quiescent background (QB) flux following the procedure in \cite{Meyer2019}. Their algorithm consists in finding an optimal distribution between the minimum and the mean flux of the light curve and then take the median of this distribution as the QB flux. We additionally estimated the standard deviation of the optimal distribution as the uncertainty. The QB flux for the gamma-rays is shown as a red dash-dotted line in Figure~\ref{multiwavelength}. We define a gamma-ray flare when the flux is outside the uncertainty range of the QB flux.

The first gamma-ray flaring event occurred around JD$_{245}=5100$ (JD$_{245}$ $\coloneqq$ JD - 2450000). This event seems to be constituted by two different flares at JD$_{245}\sim5050$ and JD$_{245}\sim5180$. There are two ejections of components from the 43 GHz core, B2 and B3 in \cite{Jorstad2017}, at JD$_{245}=4945.65\pm51.10$ and JD$_{245}=5303.35\pm98.55$, respectively. The ejections of these components are probably responsible for the two flares composing this event. This high activity is also observed in the V-band, 1~mm and 15~GHz light curves. The polarization degree during this event is not as high as during others. The apparent lack of strong variability might be an artifact of the sparse polarization observations. However, the polarization degree indicates that the synchrotron contribution to the optical emission is low or the magnetic field structure in the emission region could be inhomogeneous.

The second gamma-ray flare started at JD$_{245}\sim5500$, peaked at JD$_{245}\sim5700$ and ended at JD$_{245}=6000$. This flare in gamma-rays and V-band is structured by multiple shorter flares. It is the only major event that presents a X-ray flare. At this epochs, \cite{Jorstad2017} did not find detectable disturbances in the 43 GHz maps. Despite the lack of a component ejection, the flux from the 43 GHz radio core showed a considerable increase \citep{Jorstad2017}, indicating activity in the zone.
\cite{Raiteri2012} explained this event geometrically, due to changes in the viewing angle and therefore, variations of the Doppler factor. In Figure~\ref{multiwavelength}, we can see that the 1~mm flux shows an increase during this event, and on the contrary, the 15~GHz flux seems to decrease. Additionally, the polarization angle during this period is only varying within $\sim 90\degr$. \cite{Raiteri2012} also showed that the variations of the polarization angle during this period are smaller than during prior epochs. The polarization angle behavior implies that neither the jet structure nor the magnetic fields morphology are changing much. However, the polarization degree reaches its maximum value of the light curve showing that the optical emission from synchrotron increased during this event or is dominated by a smaller emission region with a more homogeneous magnetic field.

There are two prominent gamma-ray flares around JD$_{245}=6200$ and JD$_{245}=6500$, which coincide with ejection of components from the 43 GHz core within their uncertainties. For the first flare, we see that the ejection times of two components, overlap within their uncertainties. The B4 component ejected at JD$_{245}=6128.80\pm124.44$ and C3 at JD$_{245}=6185.5\pm30.0$ from \cite{Jorstad2017} and \cite{Algaba2018_2}, respectively. These components seem to be the same but independently detected by these research groups. The second flare coexists with the ejection of the component C2 at JD$_{245}=6520.5\pm30.0$ \citep{Algaba2018_2}. The flux from the other wavebands showed increases at this time as well. A strong flux increase in the V-band is noticeable at JD$_{245}\sim6300$ which does not have a similar counterpart in the gamma-rays, X-rays, 1~mm nor 15~GHz. However, the polarization degree shows a similar fast flare.

After these events, the gamma-rays variability decreased for about 4 years. However, there are three epochs at which the gamma-rays increased their flux above 3$\sigma$ of the QB. The gamma-ray flux increase at JD~$_{245}\sim7050$ coincides with an increase in X-rays, V-band, and the UV-continuum; the 15~GHz and 1~mm flux is very low. The gamma-ray flux increase at JD~$_{245}\sim7150$ only coincides with a peak in 15~GHz. Meanwhile, the gamma-ray flux increase at JD~$_{245}\sim7650$ is contemporaneous to a small increase of V-band and UV-continuum; the 15~GHz flux is low during this time. Additionally, there are three epochs when the gamma-rays flux surpasses the 1$\sigma$ uncertainty range of QB. At JD~$_{245}\sim6900$, the gamma-ray increase is contemporary to flux bumps in 1~mm and 15~GHz. At JD~$_{245}\sim7430$, there are increases in the continuum, the V-band, and 15~GHz, as well as an increase of the polarization degree. Finally, at JD~$_{245}\sim7870$, there are not clear connection between the gamma-ray flare with the variability of the other bands.

After JD$_{245}=8100$, the gamma-rays reached fluxes similar to previous maximums. The 1~mm and 15~GHz wavebands also exhibit strong flaring behavior. The V-band and polarization degree increase at the time, however, the data from the Steward Observatory end after JD$_{245}\sim8300$. The origins of these flares are unknown since there are no more recent imaging studies of the radio core.

\begin{figure*}[htbp]
\begin{center}
\includegraphics[width=1\textwidth]{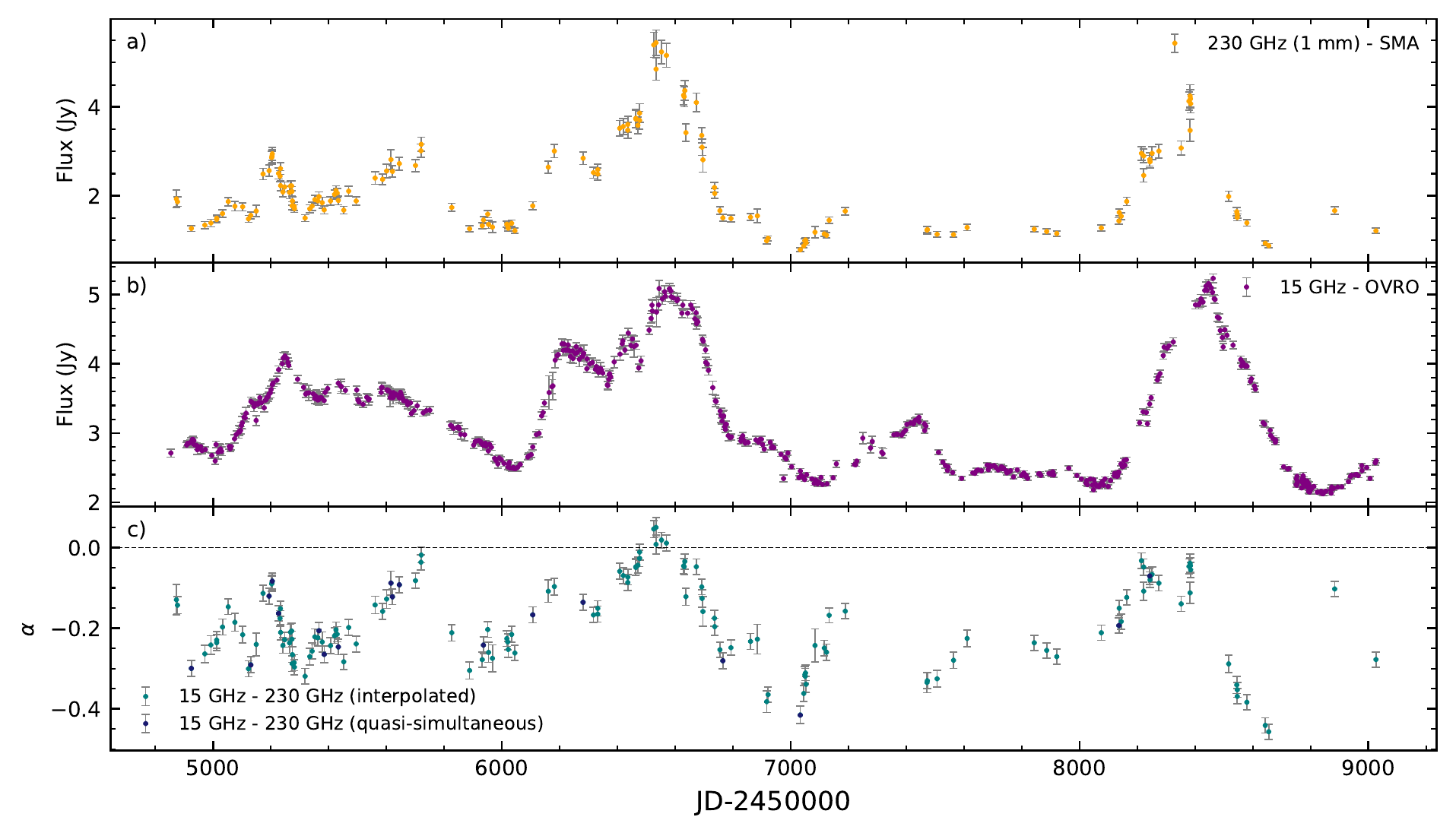}
\caption{Light curves of a) 1~mm and b) 15~GHz, for details on these see Section \ref{sec:observe}. c) The spectral index ($\alpha$, $S_\nu \propto \nu^\alpha$) estimated with the 15~GHz and 1~mm (230 GHz) flux densities. The teal points correspond to when the estimation was through interpolation and the dark blue points correspond to when there were available quasi-simultaneous epochs. The horizontal dashed line is at $\alpha = 0   $.}
\label{spectral_index}
\end{center}
\end{figure*}

The 1~mm and 15~GHz light curves are correlated with a delay of $\sim 39$ days, this can be interpreted as the 1~mm emission region being embedded in a larger 15~GHz emission region. Additionally, to extract more information about this zone, we computed the spectral index ($\alpha$, $S_\nu \propto \nu^\alpha$) with the 15~GHz and 1~mm (adopting 230~GHz as the frequency) flux densities during the quasi-simultaneous epochs, as well as, interpolating the 15~GHz light curve into the 1~mm epochs (since the 15~GHz light curve has the better sampling). The spectral index in almost the entire observation period is in the optically thin regime ($\alpha < 0$), except for a short period around JD$_{245}=5570$ where it barely surpasses the $\alpha = 0$ \citep{Fromm2011,ParkTrippe2014} separation from the optically thin and thick ($\alpha > 0$). This can be observed in Figure~\ref{spectral_index}.

In order to quantify the intensity of the variability of a light curve of length N and observations $x_i$, the variance $S^2$ could be used. However, since we want to take into account the uncertainties $\sigma_i$ of each observation, we use the fractional variability (F$_\text{var}$) instead. The fractional variability is defined \citep{Nandra1997,Edelson2002} as the square root of the normalized excess variance ($\sigma_{\text{NXS}}^2$) 

\begin{equation}
\text{F}_{\text{var}} = \sqrt{\frac{S^2 - \langle \sigma^2 \rangle}{\langle x \rangle ^2}},
\end{equation}

where $\langle \sigma^2 \rangle = \frac{1}{N} \sum_{i=1}^N \sigma_i^2$. The uncertainty was estimated as defined by \cite{Poutanen2008}

\begin{equation}
\sigma_{\text{F}_{\text{var}}} = \sqrt{\text{F}_{\text{var}}^2 + \text{err}(\sigma_{\text{NXS}}^2)} - \text{F}_{\text{var}},
\end{equation}

where

\begin{equation}
\text{err}(\sigma_{\text{NXS}}^2) = \sqrt{\left(\sqrt{\frac{2}{N}}\frac{\langle \sigma^2 \rangle}{\langle x \rangle ^2}\right)^2+\left(\sqrt{\frac{\langle \sigma^2 \rangle}{N}}\frac{2 \text{F}_{\text{var}}}{\langle x \rangle}\right)^2}
\end{equation}

is the uncertainty of the normalized excess variance, following \cite{Vaughan2003}.

In Table~\ref{Table_Fvar}, the results for each wavelength including the estimated $\lambda$1350~\AA\ continuum are displayed. 

\input{Table3_Fvar}

The results showed that gamma-rays are the most and the 15~GHz the least variable, as expected. It is worth noting that the V-band resulted to be more variable than the $\lambda$1350~\AA\ continuum and the X-rays. However, due to the poor cadence of the X-rays, the found value might not represent the actual variability of the band.

\subsection{C~IV and $\lambda$1350~\AA\ continuum flux variability} \label{ssec:linevar}

\begin{figure*}[htbp]
\begin{center}
\includegraphics[width=1\textwidth]{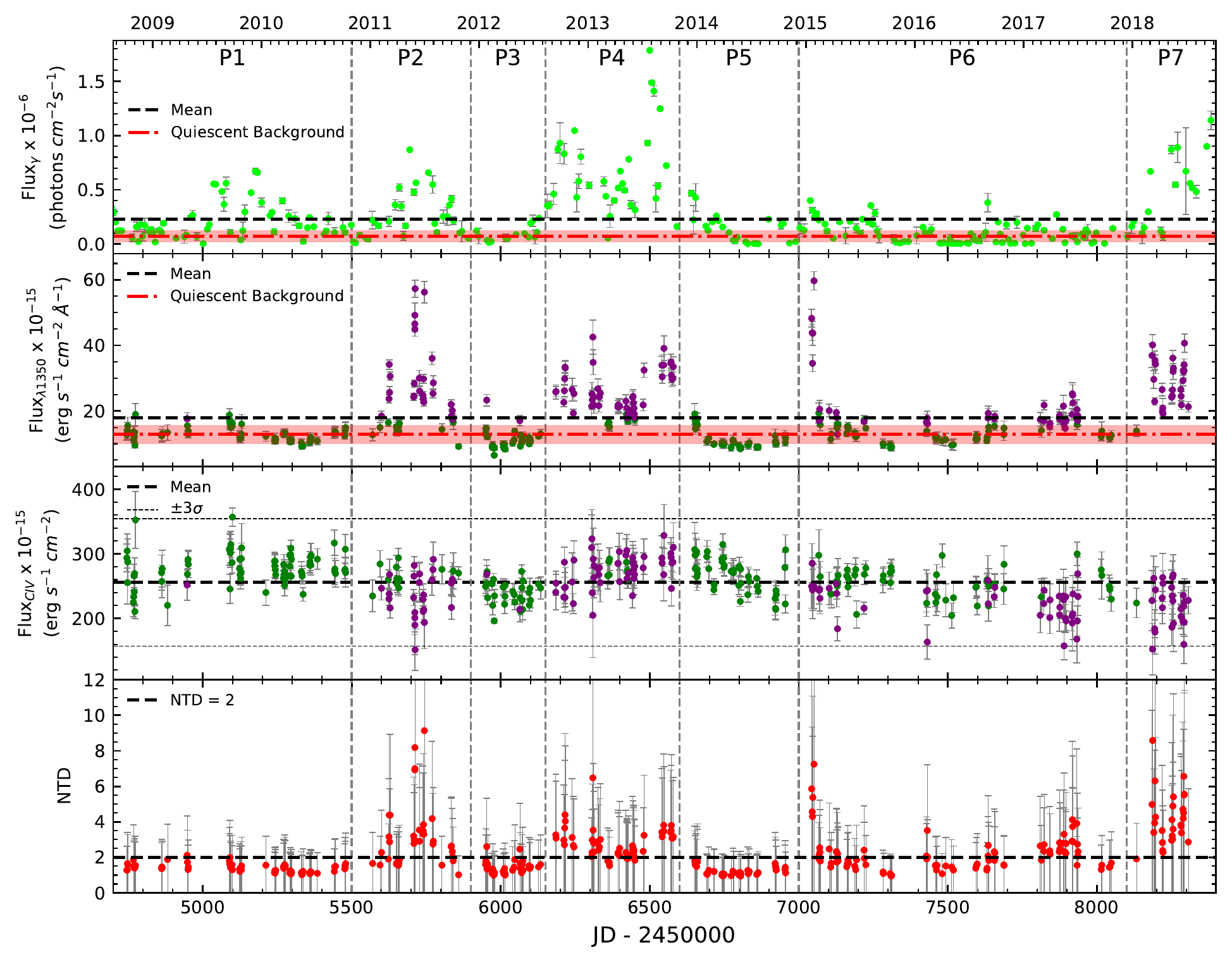}
\caption{Light curves of a) gamma-rays, b) the $\lambda$1350~\AA\ continuum flux, c) the C~IV $\lambda$1549~\AA\ emission line flux, and d) the Non-Thermal Dominance (NTD) parameter. In the middle two panels, the green points show the observations when the NTD value is below 2 and the purple points show when the NTD is above 2. The thick dashed line represents the mean of its respective light curve and the thin dashed lines represent $3\sigma$. In the bottom panel the dashed line represents NTD=2. The red dash-dotted line represents the quiescent background (see subsection \ref{ssec:linevar}). The shadowed area around the quiescent background line represents $1\sigma$. The periods defined in Section \ref{sec:var} are separated by vertical dashed lines.}
\label{line-continuum}
\end{center}
\end{figure*}

The $\lambda$1350~\AA\ continuum and C~IV $\lambda$1549~\AA\ emission-line light curves are presented in Figure~\ref{line-continuum} along with the NTD estimation for each observation. In this figure, the green color marks a dominant thermal disk contribution to the continuum emission, the purple color a dominant non-thermal jet contribution, according to the NTD.

During the major gamma-ray flaring events, we see an increase in the $\lambda$1350~\AA\ continuum flux. The continuum behaves similarly to the V-band emission as previously found through the cross-correlation analysis (see Table~\ref{Table_CC}). Meanwhile, the C~IV flux exhibits low variability (F$_\text{var} = 0.09\pm0.01$) and there is no significant correlation with the gamma-rays.

In the top panel of Figure~\ref{line-continuum}, the dominant source of continuum (disk and jet) seems to be alternating between different epochs, as shown by the NTD~$=2$ threshold with the green ($1<$~NTD~$<2$) and purple (NTD~$>2$) points. We used this to define periods in the following way: JD$_{245}=4700-5500$ (Period 1 - P1), JD$_{245}=5500-5900$ (Period 2 - P2), JD$_{245}=5900-6150$ (Period 3 - P3), JD$_{245}=6150-6600$ (Period 4 - P4), JD$_{245}=6600-7000$ (Period 5 - P5), JD$_{245}=7000-8100$ (Period 6 - P6), JD$_{245}=8100-8400$ (Period 7 - P7). Note that the P6 does not have a clear dominant source of continuum, Since, this is a different behavior, it was defined as its own period.

Since we are also expecting to have a quiescent flux and a flare-like behavior from the $\lambda$1350~\AA\ continuum flux, we estimated a QB flux using the algorithm of \cite{Meyer2019}. The QB level helps us to identify small increases in the flux which otherwise would be hard to differentiate from quiescent flux fluctuations. 

In Figure~\ref{line-continuum}, we can see that the C~IV flux deviated from the mean by approximately $3\sigma$ at times which coincide with significant increases of the continuum flux from the QB. The green points ($1<$~NTD~$<2$) at $\sim +3\sigma$ in the C~IV flux light curve coincide with changes of the continuum driven by the accretion disk. Meanwhile, the purple points (NTD~$>2$) at $\sim -3\sigma$ coincide with changes of the continuum driven by the jet. It is worth noting that these changes of $\sim3\sigma$ in the C~IV flux could be due to intrinsic variability. However, we performed a test to identify the expected number (or percentage) of observations outside the $3\sigma$ interval for light curves with the same statistical properties as that of C~IV flux. We simulated 10,000 light curves by including red noise and a power law, with the same spectral index as the observed data in the phase space following the methodologies described in \cite{Timmer1995}, \cite{Schreiber1996}, and \cite{Emmanoulopoulos2013}. For more details on this simulation method, see Appendix \ref{apB}. As a result, we found that the expected data points that exceed the 3$\sigma$ threshold are the $0.27\pm0.25$\%. For the observations, the data points that exceed the 3$\sigma$ threshold represent the $\sim$1\%, and the $\sim$4\% considering the data points with their uncertainty. These percentages are higher than expected by the simulations of the same variability profile, indicating that these points might be related to physical processes and are not statistical fluctuations.

During P1, the NTD parameter indicates that the continuum emission is dominated by the accretion disk (NTD~$<2$). We can see at JD$_{245}\sim4770$ and JD$_{245}\sim5100$ increases of $\sim 3\sigma$ from the mean C~IV flux. The changes in the continuum flux during this period are much smaller than those driven by the jet during other periods. Furthermore, there are two ejections from the radio core and there is major activity in radio, 1~mm and gamma-rays, indicating jet activity. The lack of data from the Steward Observatory in the exact dates when the gamma-ray flares occurred does not allow us to study the relationship with the continuum and the emission line during these times.

The continuum during P2 and P4 was dominated by the jet (NTD~$>2$), while during P3 and P5 by the disk (NTD~$<2$). The P6 shows a change of domination between the jet and disk. Finally, P7 is completely jet-dominated which can be confirmed with the strong flaring behavior in gamma-rays and radio frequencies.

\subsection{C~IV profile variability}

We separated the Steward Observatory spectra for each slit width setup used for the spectroscopic observations, resulting in 152 spectra obtained with 3\arcsec, 202 with 4\farcs1, and 10 with 5\farcs1. Since there are very few observations taken with 5\farcs1, these were not included in the analysis. Next, we removed the additional components and the continuum to obtain the isolated C~IV profile. Finally, we estimated the mean spectrum ($F_\lambda$) and root mean squared (RMS) spectrum

\begin{equation}
S_{\lambda} = \sqrt{\frac{1}{N} \sum\limits_{i=1}^{N} (F_{\lambda}^i - F_\lambda)^2},
\end{equation}

for each subset. Where $F_{\lambda}^i$ represents the monochromatic flux at a given $\lambda$, for the $i$th spectrum, and $N$ the total number of spectra.

\begin{figure}[htbp]
\begin{center}
\includegraphics[width=0.48\textwidth]{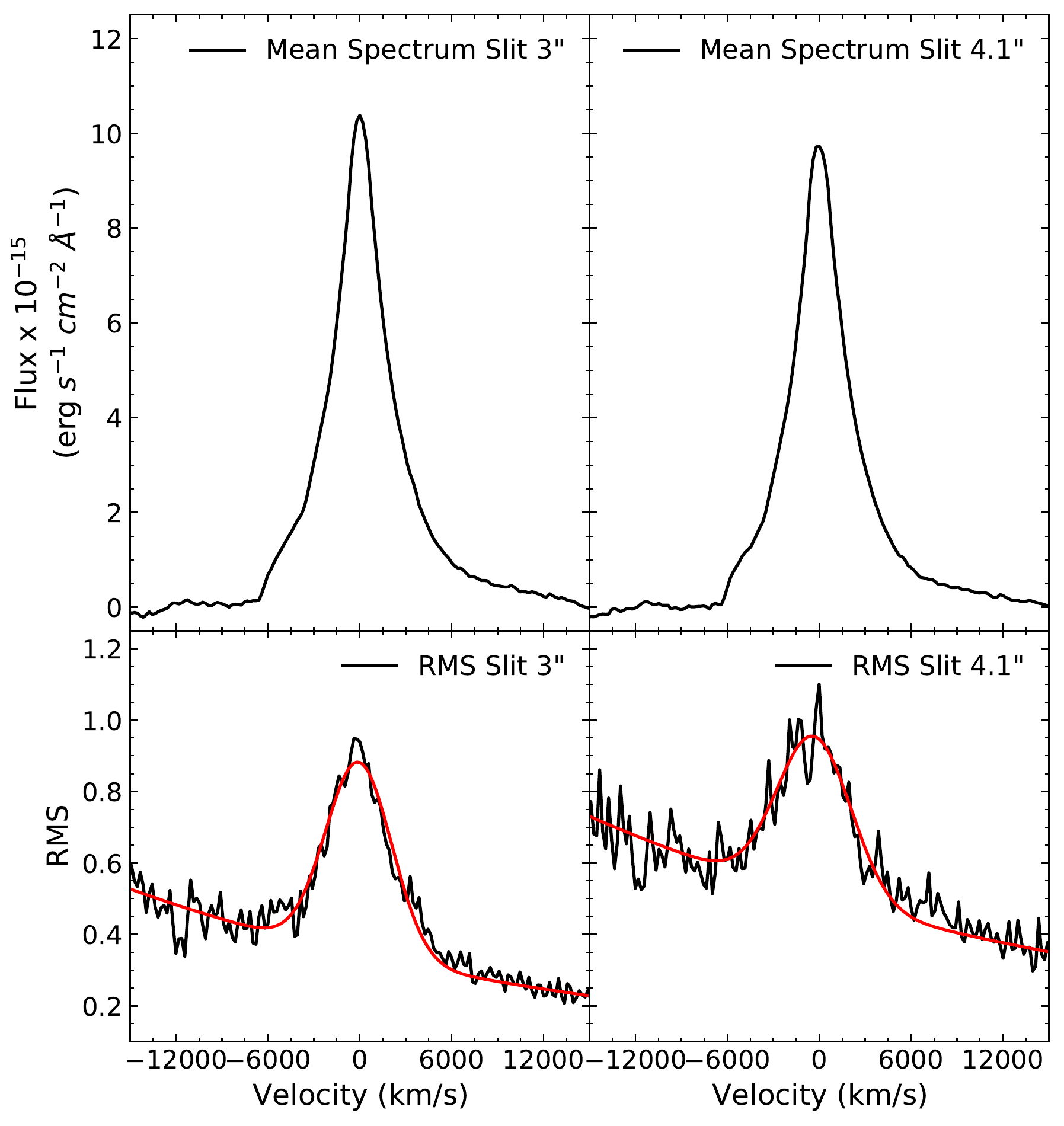}
\caption{Mean and RMS spectra of the C~IV profile for each slit width setup from the Steward Observatory. The fitted models to each RMS spectrum are represented by the red lines.}
\label{mean-rms}
\end{center}
\end{figure}

The resultant mean and RMS spectra are displayed in Figure~\ref{mean-rms}. The RMS spectra were fitted using a power-law and a gaussian component. The RMS spectra represent the variable component of the profile, thus the fit shows that the variable component has a slope, which indicates that there is more change in the blue part of the C~IV profile. Additionally, we estimated the integrated flux from the mean and RMS spectra and found that the flux from the RMS spectra corresponds to $\sim 19\%$ (3\arcsec) and $\sim 27\%$ (4\farcs1) of the flux of the mean spectra. 
Moreover, the gaussian components (RMS spectra) are slightly blue shifted. For the case of the 3\arcsec setup, the blueshift is $\sim 100$ km/s and for the 4\farcs1 setup, the blueshift is $\sim 400$ km/s. This difference in results for the different slit width setups is probably induced by a bias, since the observations tend to be taken with larger slit widths as the object gets brighter.

\section{Luminosity relations}\label{sec:LumRel}

\subsection{Luminosity relation for the full period of observations}

\begin{figure}[htbp]
\begin{center}
\includegraphics[width=0.48\textwidth]{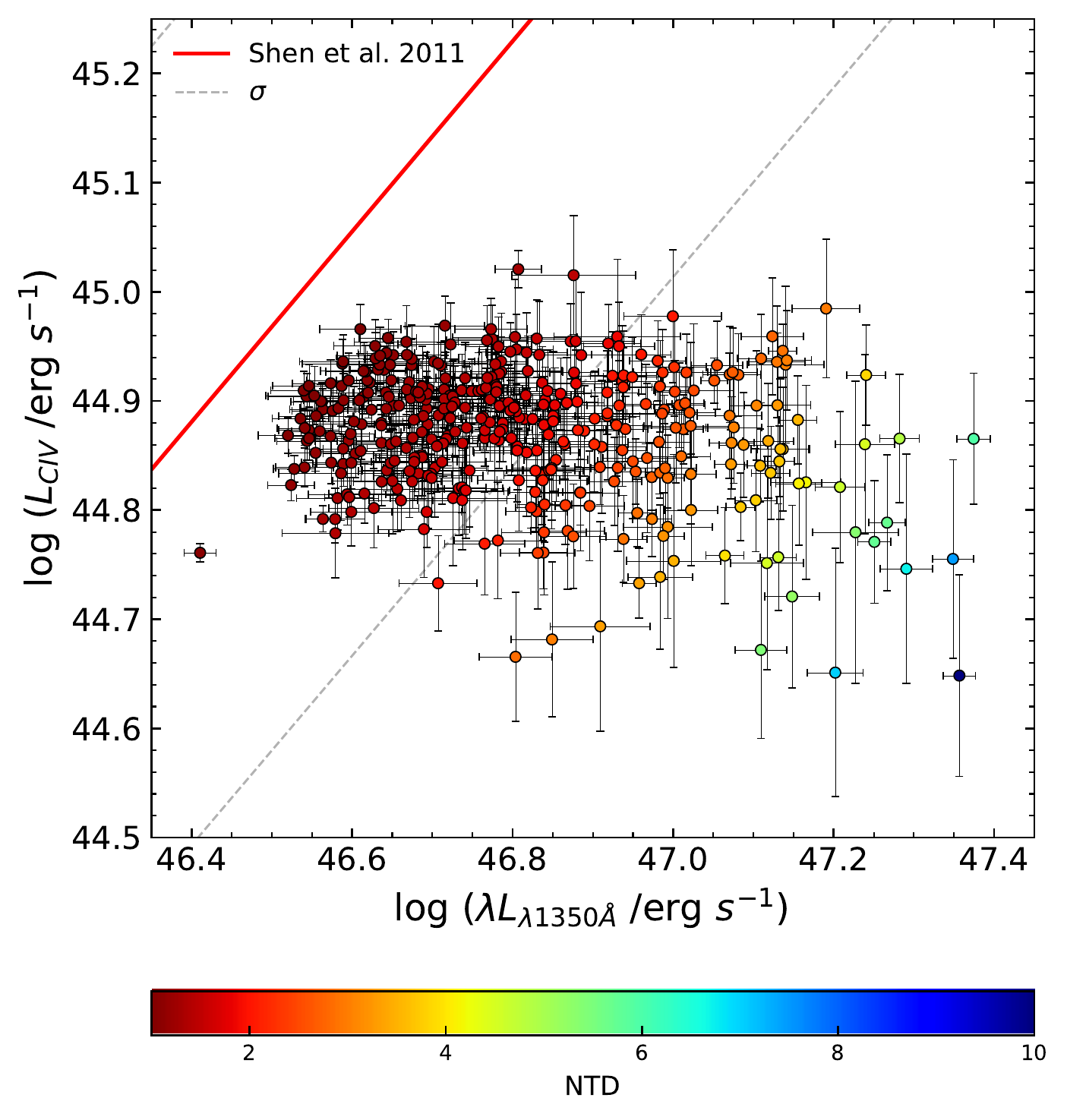}
\caption{Variation of the C~IV $\lambda$1549~\AA\ emission-line luminosity compared to the $\lambda$1350~\AA\ continuum luminosity, for B2~1633+382. The color bar indicates for each observation the non-thermal dominance (NTD) value. Disk dominance applies for NTD values between 1 and 2 while the jet dominance begins after 2. The red solid line and dashed lines represent the \cite{Shen2011} relation for a non-blazar sample and its uncertainty at $\sigma$, respectively. }
\label{ShenLine}
\end{center}
\end{figure}

Comparing the UV-continuum and the C~IV $\lambda$1549~\AA\ emission line directly can show us a different perspective of their relationship from the one given by the cross-correlations. We estimated the luminosities of the $\lambda$1350\AA\ continuum and the C~IV $\lambda$1549~\AA, and compared them to the luminosity relation estimated for a non-blazar sample by \cite{Shen2011}. Since \cite{Shen2011} performed the cosmological correction to the flux of the form $(1+z)$ to their spectra, we took this into account when estimating the luminosities. The luminosity relation is displayed in Figure~\ref{ShenLine}.

In order to determine if there was a linear relation between the luminosities of the continuum and the C~IV emission line, we performed a linear fit on the data points and their uncertainties using Orthogonal Distance Regression from the \texttt{SciPy ODR package}\footnote{\url{https://docs.scipy.org/doc/scipy/reference/odr.html}}. The resultant slope was $0.06\pm0.01$ which indicates a very flat relation, however, this result itself is not reliable since the p-value ($p_v$) was 1 to machine accuracy. Here the $p_v$ indicates the probability of obtaining a chi-square value equal or higher by chance and a value below 0.05 will be considered statistically significant. Additionally, we performed a Spearman correlation rank test (SCRT) resulting in $\rho=-0.10$ with $p_v=0.045$ (the probability of obtaining a correlation coefficient equal or higher by chance). This result shows that there is no correlation ($\rho^2=0.01$).

Moreover, we separated the sample by NTD values. This allows us to study separately when the disk (NTD $<2$) or jet (NTD $>2$) are the dominant source of continuum. In Figure~\ref{ShenLine}, the NTD value of each observation is indicated by a color bar. Same as for the full sample, we performed a linear fit and a SCRT for each sub-sample. 

The linear fit for the NTD $<2$ sample results in a slope of $0.40\pm0.02$ with a $p_v=1$ to machine accuracy, while the SCRT resulted in $\rho=0.29$ with a $p_v=8\times10^{-8}$. There is a weak, however significant correlation between the C~IV luminosity and the continuum dominated by the accretion disk.

For the NTD $>2$ sample, the resultant slope was $0.14\pm0.03$ with a $p_v=1$ to machine accuracy and the SCRT showed a $\rho=0.13$ with a $p_v=0.122$. Here, there is not a clear behavior since none of the tests resulted statistically significant.

\subsection{Luminosity relations for the time periods}\label{ssec:lum-periods}

\begin{figure*}[htbp]
\begin{center}
\includegraphics[width=1\textwidth]{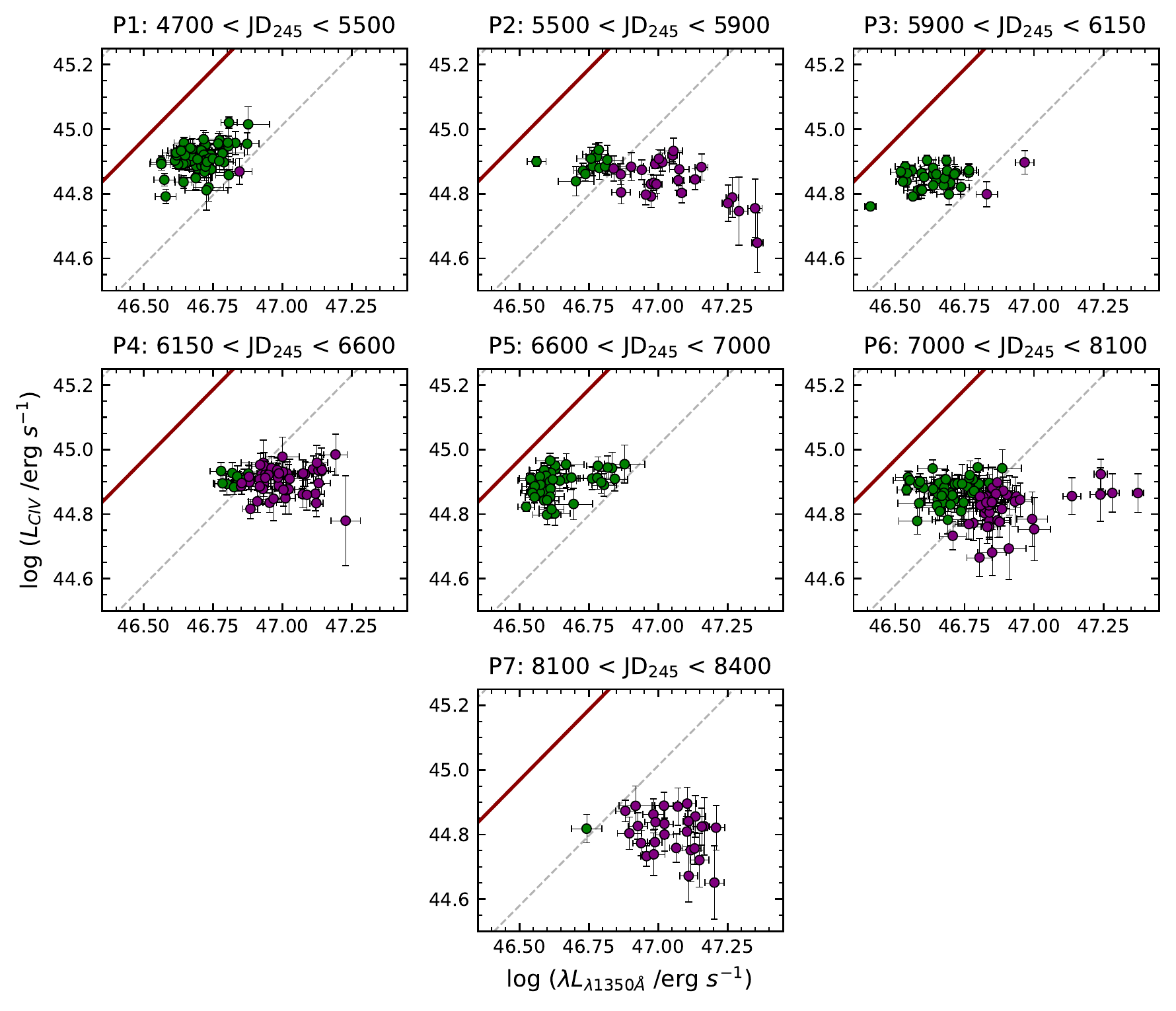}
\caption{Luminosity relation between the continuum and the C~IV emission line separated by the different periods defined in Section \ref{sec:var}. The observations are broken into the disk (NTD~$<2$) and jet (NTD~$>2$) dominance, green and purple dots, respectively. The red solid line and dashed lines represent the \cite{Shen2011} relation for a non-blazar sample and its uncertainty at $\sigma$, respectively.}
\label{lum-periods}
\end{center}
\end{figure*}

From the C~IV and continuum flux variability analysis in Section \ref{sec:var}, we saw that there were periods when the dominant source of continuum was the disk and others when it was the jet. To further investigate the relation between the continuum and the C~IV emission-line luminosities, we studied this for each of the periods previously defined in Section \ref{sec:var}. The luminosity relation for each period is shown in Figure~\ref{lum-periods}.

We performed the same analyses as for the complete sample, a linear fit and a SCRT.
The linear fits performed for these sub-samples are all statistically insignficant. However, the SCRT results showed some mixed statistical significances. We consider results statistically significant for $p_v < 0.05$.
The complete results are displayed in Table~\ref{Table_corr-periods}.

\input{Table4_corr-periods}

The P2 showed the strongest significant anti-correlation ($\rho=-0.47$) while the jet is the dominant source of continuum. Meanwhile, the P5 showed the strongest significant correlation ($\rho=0.37$), but in this case, the accretion disk continuum dominates.

\section{Discussion} \label{sec:disc}

We have studied the light curves of the different photometric and spectral features through cross-correlation and variability analyses. In addition to this, we studied the relation between the $\lambda$1350~\AA\ continuum and the C~IV $\lambda$1549~\AA\ emission-line luminosities. In this section, we will be discussing the most important results.

\subsection{The gamma-ray emission-region location}

Through the cross-correlation analysis, we found a delay of $-69.5\pm8.7$ days between the 15~GHz and the gamma-rays. This is compatible with the results found by \cite{Algaba2018_1} and \cite{Wang2021}, which they used to calculate a separation between the 15~GHz and gamma-ray emission regions of $40\pm13$ pc and $14^{+0}_{-2.4}$ pc, respectively. Additionally, using the location suggested by \cite{Pushkarev2012} of the 15~GHz core ($\sim 41$ pc) from the jet apex, \cite{Algaba2018_1} and \cite{Wang2021} estimated the location of the gamma-ray emission region as being $1\pm13$ pc and $26.5$ pc from the jet apex, respectively. We performed the same distance calculation using 

\begin{equation}\label{eq-fuhrmann}
d = \frac{\beta_{\text{app}}\ c\ \tau_{\text{source}}}{\sin{\theta}}
\end{equation}

from \cite{Fuhrmann2014}, where $\beta_{\text{app}}$ is the apparent jet speed, $c$ is the speed of light, $\tau_{\text{source}}$ is the source frame time delay, and $\theta$ is the jet viewing angle. We considered two different values of the apparent jet speed. The first is $\beta_{\text{app}} = 29.3\pm1.3$, this is estimated by \cite{Lister2013} for a single component being observed approximately three years after its ejection from the 15~GHz core in 1998. This is the value used by \cite{Algaba2018_1} and \cite{Wang2021} for their calculation. This value does not represent the jet apparent speed very well since it was estimated from a single component and from an epoch out of the observation period used for the cross-correlation analysis. This is why we estimated a second value from the more recent observations of \cite{Jorstad2017}. We took the weighted mean of the $\beta_{\text{app}}$ values for the moving components and we used the standard deviation as the uncertainty. Finally, we found the value of $\beta_{\text{app}} = 6.8\pm1.8$. For the viewing angle, the reported values are within $1\degr - 3\degr$ \citep{Hovatta2009,Liu2010}. We performed a Monte Carlo estimation of the distance between the gamma-ray and 15~GHz emission regions with parameter distributions of 10,000 values. We produced a random Gaussian distribution for each parameter except the jet viewing angle, for which we took a uniform distribution between the range of its values. The distance found using $\beta_{\text{app}} = 29.3\pm1.3$ was $d=19.1\pm6.4$ pc while using $\beta_{\text{app}} = 6.8\pm1.8$ we found $d = 4.4\pm1.9$ pc. The first value differs significantly from the value estimated by \cite{Algaba2018_1}. It is not clear from \cite{Algaba2018_1} publication whether or not they performed the cosmological correction to the time delay of the form $\tau_{\text{observed}} = \tau_{\text{source}}(1+z)$, which is required by equation (\ref{eq-fuhrmann}). From our result, the location of the gamma-rays region must be farther from the base of the jet, in agreement with the result of \cite{Wang2021}. Using the $\sim 41$ pc location of the 15~GHz core from the jet apex \citep{Pushkarev2012} and our estimations, the gamma-ray emission region must be between 22 and 37 pc from the base of the jet. However, we expect the location to be closer to $\sim37$ pc due to the more recent apparent speed measurement used.

We also found a delay of $19.3\pm16.3$ days between the gamma-rays and 1~mm. This result along with the delay found between the 15~GHz and gamma-rays would locate the 1~mm emission region between the gamma-rays and 15~GHz emission regions. This is consistent with the shock-in-jet model \citep[e.g.][]{MarscherGear1985}. In which a shock wave propagates along the jet accelerating electrons which will lose energy while emitting synchrotron radiation.

Additionally, there was no correlation at any delay between the continuum and the C~IV. However, when we separated the light curves in the moments when the continuum was either dominated by the disk (NTD $<2$) or the jet (NTD $>2$), correlations were found.
For the D-D continuum, we found a delay consistent with zero which indicates that the BLR zone which is emitting C~IV is very small. However, we do not have enough resolution to estimate an accurate value. This is consistent with the stratification of the BLR found through reverberation mapping \citep[e.g.][]{Krolik1991} and microlensing \citep[e.g.][]{Guerras2013} studies in which the high-ionization lines are emitted in a more compact zone and closer to the accretion disk than low-ionization lines.
Meanwhile, for the J-D continuum, the delay found was larger ($-227.4\pm12.9$ days). This correlation could indicate a connection between BLR material and the jet. 
Another scenario, with the accretion disk being the only ionizing source of the BLR material, is if the high activity of the disk might perturb the magnetic fields and the material close to the base of the jet inducing an increase of the continuum dominated by the jet. Hence, this scenario could indicate that the disk affects the jet.

\subsection{The origin of the gamma-ray flares}

In Section \ref{sec:var}, we found that the different bands showed different degree of variability, with gamma-rays being the most and the 15~GHz the least variable.

During P1, there is a gamma-ray flaring event. \cite{Jorstad2017} report the ejection of two components from the 43 GHz core at the time. They also found an increase in the 43~GHz core and 37~GHz light curves at the times of the ejections. We found that the 1~mm and 15~GHz show an increase in their flux during this period with 1~mm leading 15~GHz. The gamma-ray flaring event during this period was most likely caused by the radio core ejections from the perspective of the shock-in-jet model \citep{MarscherGear1985}.

The origin of the gamma-ray flare during P2 is not clear. \cite{Jorstad2017} did not find any detectable disturbances in the 43 GHz maps, however, the 43 GHz core increased its flux during this event. \cite{Raiteri2012} explained this gamma-ray flare with changes in the Doppler factor derived from geometrical effects. However, \cite{Algaba2018_2} and \cite{HagenThorn2019} found in their analyses that variations in the Doppler factor are not enough to explain the flux variations. Additionally, we found that during this period the polarization angle changed only within 90$\degr$. The small change of polarization angle during this period tells us that the morphology of the magnetic field and the structure of the jet are not changing much \citep{Marscher2008,Chandra2015,Blinov2018}. We will discuss a possible explanation for this flare later in the section.

The P3 and P5 present very low gamma-ray emission which coincides with the flux drop in the 1~mm and 15~GHz bands. This implies that the jet activity during these periods was low.

During P4, the ejection of two components from the 43 GHz core \citep{Jorstad2017,Algaba2018_2} fires up the gamma-rays, as well as, the 1~mm and 15~GHz. The spectral index of 1~mm and 15~GHz reaches positive values during the larger flare. This means that at least during the second gamma-ray flare in P4 the medium was optically thick to 1~mm and 15~GHz. This shows that the physical conditions in the emission regions during the flares were different during this event.

During P6 and P7, the mechanisms that caused the gamma-ray flares are unknown since there are not recent imaging studies of the radio core.

The $\lambda$1350~\AA\ continuum is significantly correlated to the V-band, meanwhile, the C~IV emission line flux did not show strong variability. However, Figure~\ref{line-continuum} shows that there are episodes in which the C~IV flux deviates from the mean by approximately $3\sigma$. During P1, the C~IV flux increases above the mean value significantly, which can be linked to high activity in the disk since the NTD showed thermal dominance at this period. A similar flare-like behavior of the Mg~II $\lambda 2798$~\AA\ emission line was observed for other FSRQ sources \citep{Leon-Tavares2013,Chavushyan2020}, however, in those cases the ionizing continuum was shown to be related to the jet. During the P2, P6 and, P7, there are epochs when the C~IV flux decreases below the mean value significantly and coincides with high continuum emission dominated by non-thermal processes, as well as gamma-ray flares. Here again, we see a connection between the non-thermal continuum emission and the C~IV flux. This behavior was also observed in 3C 454.3 but for the Mg~II $\lambda 2798$~\AA\ emission line \citep{Nalewajko2019,AmayaAlmazan2021}. This type of behavior is explained by \cite{Nalewajko2019} with destructive ionization, however for a high-ionization line like C~IV, this scenario is less likely than for the low-ionization line Mg~II.

To further understand the physical processes occurring at the BLR clouds, we performed the C~IV profile analysis displayed in Figure~\ref{mean-rms}. The RMS isolates the variable component of the emission line. It presented a negative slope and it is slightly blueshifted. These characteristics show that the bigger changes of the C~IV profile are happening in the blue wing. This can be interpreted as an outflow from the accretion disk carrying BLR material towards the observer (jet flow direction) causing more variability in the blue wing of the emission line. Meanwhile, if there is a similar wind on the other side of the central engine we may not see the red wing being as variable due to it being obscured by the accretion disk.

\subsection{The luminosity relation between the $\lambda1350$~\AA\ continuum and the $\lambda1549$~\AA\ C~IV emission line}

The luminosity relation between the continuum and C~IV found for our data did not show a significant correlation. Additionally, it does not completely follow the \cite{Shen2011} relation found for a quasar sample. This behavior is similar to what was found for CTA 102 \citep{Chavushyan2020} and 3C 454.3 \citep{AmayaAlmazan2021}. For PKS 1510-089 \citep{Rakshit2020}, a similar result was found but for a relation between the $\lambda5100$~\AA\ continuum and the H$\beta$ emission line of \cite{Rakshit2020_1} estimated for another quasar sample. 

However, the values within the $1\sigma$ uncertainty of the \cite{Shen2011} relation correspond to the D-D regime, which shows lesser influence of the radiation from the jet over the emission line. Therefore, the estimation of black hole mass through reverberation mapping or single-epoch techniques, should be performed using only observations when NTD~$<2$ \citep{Chavushyan2020,AmayaAlmazan2021}. However, it needs to be considered that at least part of the BLR does not seem to be in a virialized state which could affect the obtained value.
Additionally, in Figure~\ref{ShenLine}, we can see that the observations with NTD~$>2$ seem more spread out in continuum luminosity, compared to the points with NTD~$<2$. This is a possible consequence of how the relativistic beaming affects the continuum emission.

Moreover, we separated the luminosity relations into the defined periods in Section \ref{sec:var}, as shown in Figure~\ref{lum-periods}. After performing the analyses described in subsection \ref{ssec:lum-periods}, the P2 showed a significant anti-correlation while the jet was the dominant source of continuum. This shows that while the non-thermal continuum from the jet was increasing, the C~IV emission-line luminosity decreased. The aforementioned phenomenon is also present during P6 and P7 but in a weaker anti-correlation, and for the latter, not statistically significant.

\subsection{Magnetic reconnection as an alternative mechanism for the production of gamma-ray flare of 2011}

The origin of the gamma-ray flare that extends throughout 2011 has not reached a consensus. It is a strong flare present on multiple frequencies, as well as, on the optical spectroscopic features studied in this work. Here, we propose magnetic reconnection \citep[e.g.][]{Vasyliunas1975,Begelman1998} as an alternative mechanism for the production of the gamma-ray flare of 2011, which is composed by multiple shorter flares. The magnetic field lines need to get closer to create a thermal pressure gradient and accelerate particles at the base of the jet. The magnetohydrodynamic (MHD) instabilities can be described by a perturbation of the form $\propto e^{im\theta}e^{i(kz -\omega t)}$ (cylindrical coordinates). The first factor allows to have perturbations on the axis of the magnetic field, the main two modes correspond to when $m=0$ and $m=1$, named pinch and kink instabilities, respectively. In the second factor, $k$ is the wavenumber and $\omega$ the frequency. For further details see \cite{Begelman1998}.

\cite{Shukla2020} summoned the kink instability to explain a gamma-ray flare from 3C 279, in addition to the helical trajectories of VLBI components due to the non-axisymmetric nature of the perturbation.
Since the polarization angle is not changing much (within 90$\degr$) during P2, the pinch instability (axisymmetric) should be the main mode of the perturbation responsible for the magnetic reconnection in this period. Therefore, the instability should happen without a major change in the morphology of the magnetic field (to our perspective). Furthermore, \cite{HaochengZhang2020} found that kink instabilities or turbulence are unlikely to produce any large ($\geq90\degr$) polarization angle rotations since these perturbations are improbable to form nearly perfect antiparallel magnetic field lines.

\subsection{The virialized BLR and the BLR material close to the jet}

The results discussed above allow us to picture an outflow in the direction of the jet from the BLR. When the jet augments its activity, it steals BLR material through the increasing magnetic fields from the virialized clouds \citep[e.g.][]{PaltaniTurler2003}. \cite{Chatterjee2019} found with general relativistic MHD simulations that the interaction between the jet and wind near the base of the jet induces oscillations which drive pinch instabilities. They explain that these instabilities also alter the jet dynamics with the mass transferring into the jet, and this might help form the inner fast and slow layer known as the spine-sheath structure \citep[e.g.][]{Ghisellini2005}.

Under aforementioned scenario, the C~IV flux decreases when the continuum is J-D since the available material to become ionized by the disk would have diminished. The dragged material is not ionized by the jet as efficiently as the material from the virialized BLR by the disk, since the density of the former could be significantly smaller than the one of the latter; the density being a key parameter for the cloud emissivity \citep[e.g.][]{Emmering1992,Peterson2006}. Furthermore, there is probably not enough material accumulated close to the radio core to form an additional BLR to be easily ionized by the non-thermal continuum and induce an increase in the emission line flux, as seen in CTA 102 \citep{Chavushyan2020} and 3C 454.3 \citep{Leon-Tavares2013,AmayaAlmazan2021} for the Mg~II $\lambda 2798$~\AA\ emission line.

\section{Summary} \label{sec:sum}

The blazar B2~1633+382 ($z=1.814$), also known as 4C~38.41, has exhibited multiwavelength variability, as well as, flaring periods of high energy emission over the years. In this study, we analyzed the optical spectroscopic data (2008-2019) from the ground-based observational support of the Fermi Gamma-ray Space Telescope at the University of Arizona. From these optical spectra, we obtained the prominent C~IV $\lambda$1549~\AA\ emission line and the UV-continuum. These spectral features are compared and analyzed alongside other photometric observations of multiple wave bands freely available on the web. Through different analyses we found the following results:

\begin{itemize}

\item We found a distance between the 15~GHz and the gamma-ray emission regions of $4.4\pm1.9$ pc. From the value found in the literature of the location of the 15~GHz core ($\sim 41$ pc from the base of the jet), the gamma-ray emission region might be located at $\sim 37$ pc from the jet apex. 

\item There is no correlation at any delay between the $\lambda$1350~\AA\ continuum and the C~IV $\lambda$1549~\AA\ emission line. However, we separated the observations at the times when the disk or the jet were the dominant sources of continuum and correlations appeared. When the disk was dominating, we found a significant correlation at a delay consistent with zero, which indicates that the C~IV emission zone in the BLR is very compact. When the jet was dominating, we found a larger delay, which might indicate a connection between the jet and BLR material or that the disk is affecting the jet (in a scenario with the disk as the only ionizing source).

\item On the one hand, during significant increases of disk dominated continuum from the quiescent background flux, the C~IV flux increased by $\sim3\sigma$ from the mean. On the other hand, during increases of the jet dominated continuum, the C~IV flux decreased $\sim3\sigma$ from the mean.

\item There is a larger variable component in the blue wing of the C~IV emission-line profile. This can be interpreted as an outflow of BLR material in the direction of the jet.

\item The luminosity relation between the continuum and the C~IV of our data does not follow the one for a quasar sample dominated by radio-quiet sources. However, the observations when the disk is the dominant source of continuum lie within 1$\sigma$ of this relation.

\item The luminosity relation between the continuum and the C~IV was separated into time periods defined by the dominant source of continuum (disk or jet). This showed a statistically significant anti-correlation during the flare of 2011, when the jet was dominating the continuum. This means that the luminosity of C~IV emission diminished as the continuum increased.

\end{itemize}

These results lead us to propose that during the flaring period of 2011, since there are no ejections from the radio core reported at the time, the gamma-ray flare is produced by a magnetic reconnection which causes acceleration of particles. This magnetic reconnection is driven by a MHD instability, which can be produced by interactions between the jet and wind being dragged by the jet. This also explaining the decrease of C~IV flux during this period. Finally, B2 1633+382 does not have much BLR material close to the jet or this material is not dense enough to be ionized by the non-thermal continuum.

\acknowledgments 

Acknowledgments. We thank the anonymous referee for the constructive comments that helped to improve the manuscript. This work was supported by CONACyT (Consejo Nacional de Ciencia y Tecnolog\'ia) research grant 280789 (M\'exico). RAA-A acknowledges support from the CONACyT program for PhD studies. This work is supported by the MPIfR-Mexico Max Planck Partner Group led by VMP-A. Data from the Steward Observatory spectropolarimetric monitoring project were used. This program is supported by Fermi Guest Investigator grants NNX08AW56G, NNX09AU10G, NNX12AO93G, and NNX15AU81G \url{http://james.as.arizona.edu/~psmith/Fermi/}. The CSS survey is funded by the National Aeronautics and Space
Administration under Grant No. NNG05GF22G issued through the Science
Mission Directorate Near-Earth Objects Observations Program.  The CRTS
survey is supported by the U.S.~National Science Foundation under
grants AST-0909182. The 1~mm flux density light curve data from the Submillimeter Array was provided by Mark A. Gurwell. The Submillimeter Array is a joint project between the Smithsonian Astrophysical Observatory and the Academia Sinica Institute of Astronomy and Astrophysics and is funded by the Smithsonian Institution and the Academia Sinica. This research has made use of data from the OVRO 40-m monitoring program (Richards, J. L. et al. 2011, ApJS, 194, 29), supported by private funding from the California Insitute of Technology and the Max Planck Institute for Radio Astronomy, and by NASA grants NNX08AW31G, NNX11A043G, and NNX14AQ89G and NSF grants AST-0808050 and AST- 1109911. \\

\software{Astropy \citep{astropy2013, astropy2018},
FTOOLS \citep{Blackburn1995},
XSPEC  \citep{Arnaud1996}, 
Fermitools (v 1.0.20), SciPy \citep{2020SciPy-NMeth}
}

\appendix 

\section{Cross-correlation analysis figures}\label{appendix}

\begin{figure*}[ht!]
\begin{center}
\includegraphics[width=0.96\textwidth]{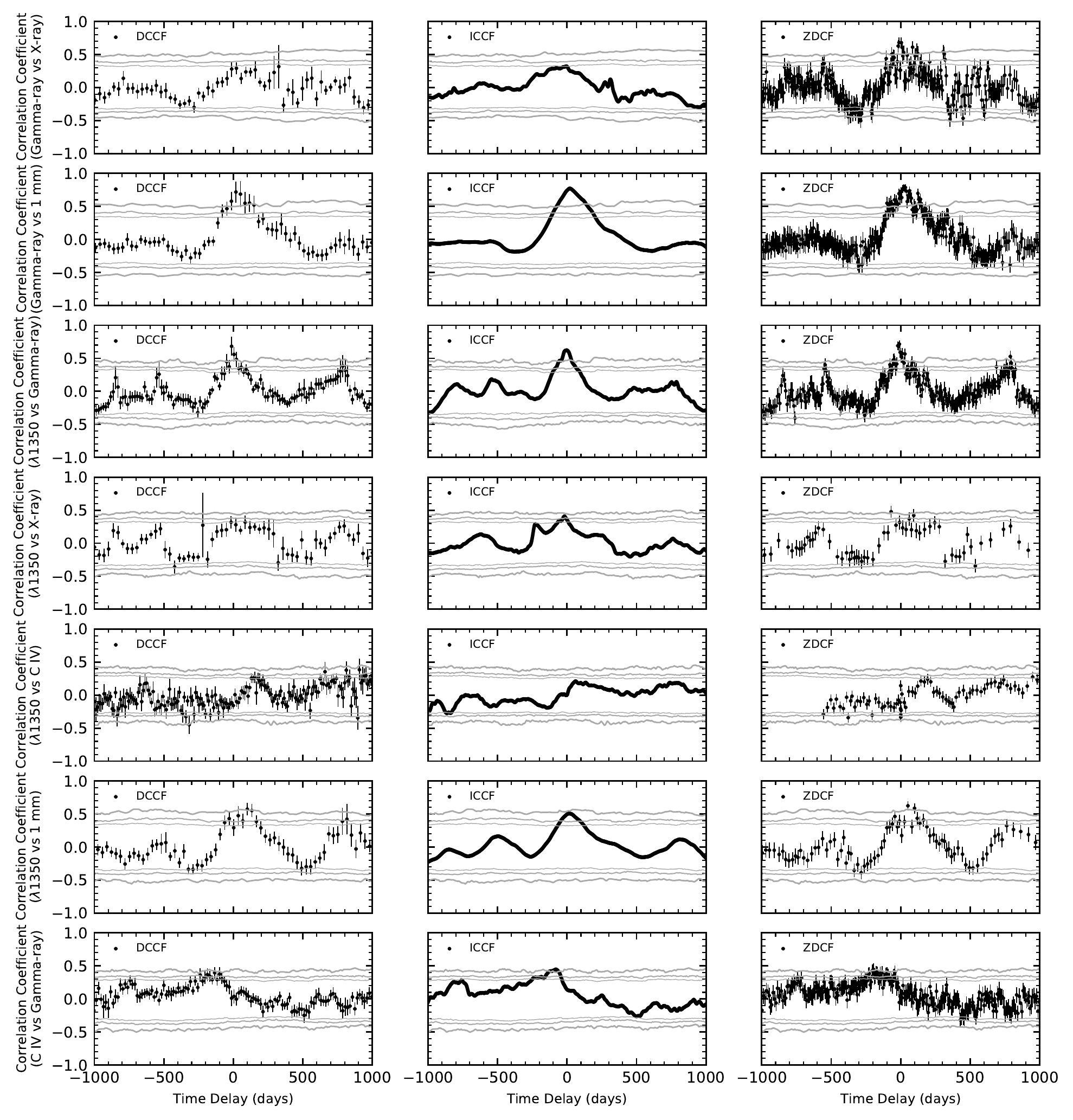}
\caption{Cross-correlations (see Section \ref{sec:crosscorr}). Each row corresponds to an analysis done for the two respective features using three different methods. The first method is the Discrete Cross-Correlation Function (DCCF), the second is the Interpolation method (ICCF), and the third method is the Z-transformed Cross-Correlation Function (ZDCF). The significance at the 90\%, 95\%, and 99\% are represented by grey lines. The time delays that were found not to be aliases are listed in Table~\ref{Table_CC}.}
\label{cc-fig1}
\end{center}
\end{figure*}

\begin{figure*}[ht!]
\begin{center}
\includegraphics[width=0.96\textwidth]{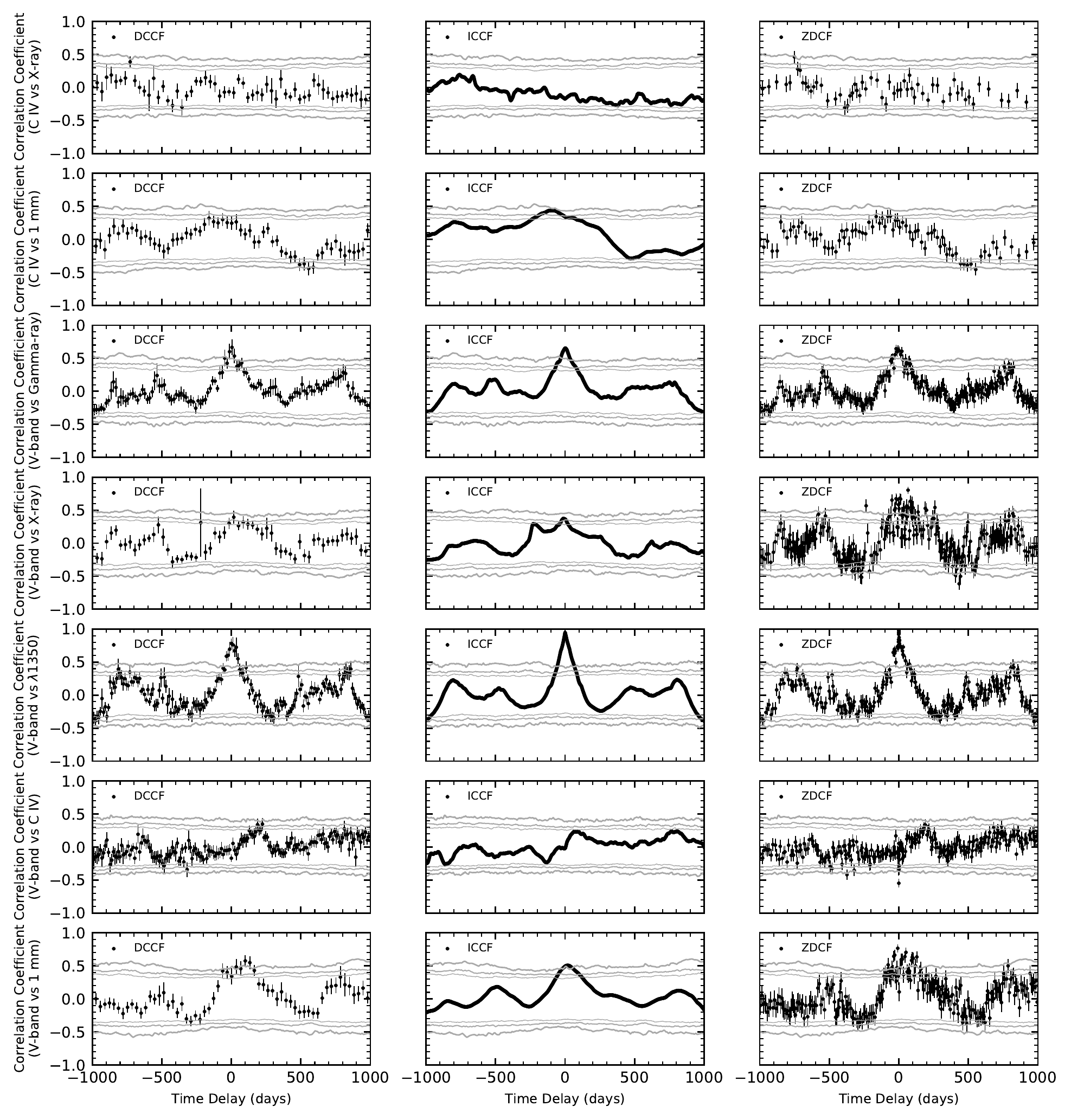}
\caption{Cross-correlations (see Section \ref{sec:crosscorr}). The lines represent the same as in Figure~\ref{cc-fig1}. The time delays that were found not to be aliases are listed in Table~\ref{Table_CC}.}
\label{cc-fig2}
\end{center}
\end{figure*}

\begin{figure*}[ht!]
\begin{center}
\includegraphics[width=0.96\textwidth]{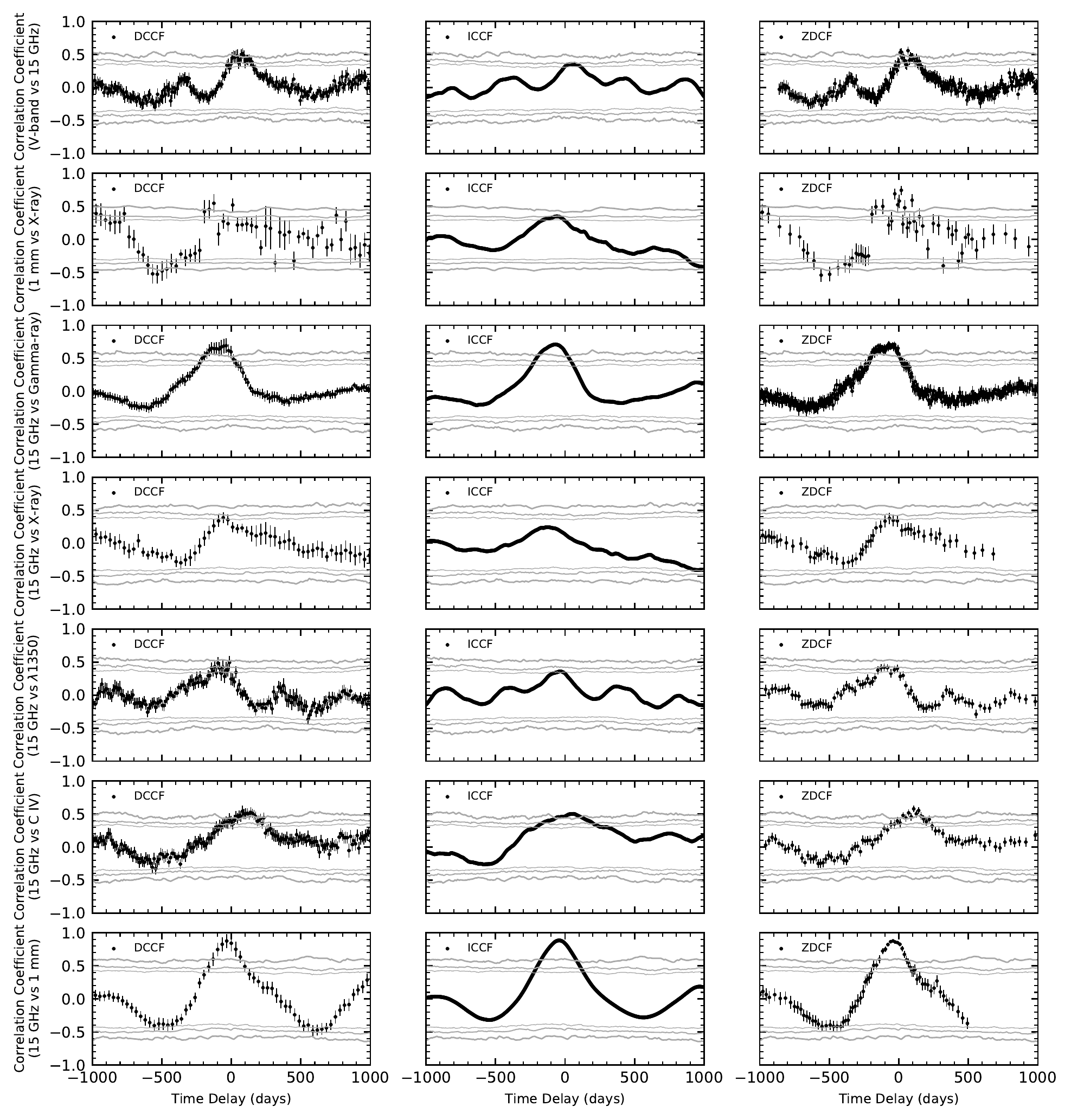}
\caption{Cross-correlations (see Section \ref{sec:crosscorr}). The lines represent the same as in Figure~\ref{cc-fig1}. The time delays that were found not to be aliases are listed in Table~\ref{Table_CC}.}
\label{cc-fig3}
\end{center}
\end{figure*}

\FloatBarrier

\section{Cross-Correlation Analysis}\label{apB}

The Cross-Correlation function has been used in science for more than 50 years. It is a very useful tool for analyzing time series. A specific application in astronomy is to quantify lags between different types of emission in Active Galactic Nuclei (e.g. between continuum and emission-line flux variations, see \citealp{Gaskell1986,Gaskell1987}).

In this work we used three different methodologies of the cross-correlation function: the Interpolated Cross-Correlation Function of \cite{Gaskell1986}, the Discrete Cross-Correlation function of \cite{Edelson1988} and the Z-Transformed Discrete Cross-Correlation Function of \cite{Alexander1997}; in order to quantify the lags between the different parts of the electromagnetic spectrum. Some of the details presented here can be found in \cite{PatinoAlvarez2013}.

\subsection{Interpolated Cross-Correlation Function}

To use the Interpolated Cross-Correlation Function (ICCF) method, we need the data from both light curves to be taken uniformly (i.e. data in the same moment for both curves and with equal spacing). Given that this is extremely difficult to do in real life, the alternative that \cite{Gaskell1986} suggest is to make linear interpolation in the data, so we can get two vectors of the same size, with fluxes in the same moment and observation times equally spaced.

Linear interpolation is used because it will lesser the effect of any assumed variability due to the interpolation process (compared to, e.g., cubic spline, Chebyshev, Lagrange, etc.). In practice, if the light curves do not have the exact same time-frame (which is very common when dealing with multiwavelength light curves), then we should take into account, that we just want to do interpolation, and not extrapolations; since these can lead to very large fluxes, or unphysical negative fluxes. To avoid this, we take as starting point of the cross-correlation analysis, the first point of the light curve that starts later; and as ending point, the final point of the curve that finishes earlier.

In the original paper of \cite{Gaskell1986}, they used this method to look for correlations between continuum and line flux measurements, and one particular step they did was to extend the continuum measurement backwards and forward: they added fluxes before the starting point, with the same value as the first point, and added fluxes after the ending point, with the same value as the last point; according to their work, this minimizes the introduction of spurious correlations. This was done because of the very scarce sampling they had. Given that the sampling in the light curves used in this work is considerably better, we do not need to do this. Also, even when the interpolation method works by assuming a certain understanding of the behavior of the light curve, applying this step might in fact cause the opposite effect the authors were looking to avoid (see also \citealp{White1994}).

Once we have two flux vectors (from different light curves) covering a time interval $T$, we generate a vector of lags, that can have values from $-T$ to $+T$. However, this is exact only when we have the interpolated fluxes in the light curves spaced by one unit of time (e.g. days). In the case when the interpolation interval is different from one unit of time, say $\Delta t$, the vector of lags should have values

\begin{equation}
L_i=\frac{l_i}{\Delta t},
\end{equation}

where $l_i$ is the real lag (in days for this case), in the same units as $\Delta t$; and $L_i$ must always be an integer.

This method basically moves the time axis of one of the light curves to greater or lower values, according to the lag; then, between the time interval when both light curves intersect, it computes the correlation coefficient between these segments of the light curves.

Something that must be taken into account is that if we make our lags $l_i$ go from $-T$ to $+T$, both of these lags would be computed with just one point in each light curve (because just one point will coincide once the light curve has been displaced that much), and lags close to these values will also have just a few points. In this situation a correlation coefficient is not statistically significant. In this work, we took as minimum $l_i=-0.8T$, and as maximum $l_i=+0.8T$ with the purpose of leaving at least 20\% of the light curves to obtain a correlation coefficient.

Now, if we have two flux vectors $\mathbi{x}$ and $\mathbi{y}$, and a vector of lags $\mathbi{L}$, we can calculate the Cross-Correlation Function $P_{xy}(L)$ by using

\begin{equation}
P_{xy}(L) = \left\{ \begin{array}{ll}
\frac{\displaystyle
\sum_{k=0}^{N-|L|-1}(x_{k+|L|}-\bar{x})(y_k-\bar{y}) }
{\displaystyle
\sqrt{ \biggl[\sum_{k=0}^{N-1}(x_k-\bar{x})^2\biggr]
\biggl[\sum_{k=0}^{N-1}(y_k-\bar{y})^2\biggr] } } & For\ L<0\\

\frac{\displaystyle
\sum_{k=0}^{N-L-1}(x_k-\bar{x})(y_{k+L}-\bar{y}) }
{\displaystyle
\sqrt{ \biggl[\sum_{k=0}^{N-1}(x_k-\bar{x})^2\biggr]
\biggl[\sum_{k=0}^{N-1}(y_k-\bar{y})^2\biggr] } } & For\ L\ge0\\
\label{corrint}
\end{array} \right.
\end{equation}

defined by \cite{Fuller1976}, where $\bar{x}$ is the mean of the points in the $\mathbi{x}$ curve that are involved in the calculation of the correlation coefficient for each lag, and similarly for $\bar{y}$.

In some works found in the literature (e.g. \citealp{Gaskell1986,Gaskell1987}), to perform the Cross-Correlation Analysis, they take $\bar{x}$ as the mean of the entire light curve, so one does not have to recalculate this value for each lag. According to \cite{White1994}, however, this is only valid, if we assume that the light curve is stationary, i.e., the mean and standard deviation do not change with time; but this is something we cannot assume for AGN light curves, and they propose to take into account for the computation of the mean, just the points involved in the correlation for each lag.

Once we are done with this step, then to represent graphically the Cross-Correlation Function, we should plot $\mathbi{L}$ $\times$ $\Delta t$ vs. $P_{xy}(L)$.

It should be taken into account that the interpolation method can yield results even from a very limited sampling, however if the interpolation is not a reasonable approximation of the real light curve, the results can be uncertain.

\subsection{Discrete Cross-Correlation Function}

The method proposed by \cite{Edelson1988}, unlike the interpolation method, does not assume any understanding of the real light-curve behavior; it uses only real data points.

The correlation function itself is binned on time intervals $\delta t$, so that the value of the DCCF at a lag $\tau$ is the average over the interval $\tau-\delta t/2$ to $\tau+\delta t/2$. In this work, we used $\delta t=sm\times1.2$, where $sm$ is the largest sampling mean of both curves. This is because, making the bin a little larger than the sampling mean will assure we have enough points in the bin for good statistics. Following is the explanation of the method.

If we have two vectors $\mathbi{a}$ and $\mathbi{b}$, then we can form pairs $(a_i,b_j)$, each one of these is associated with the pairwise lag $\Delta t_{ij}=t_j-t_i$. Now, for each of these pairs, we can compute the unbinned discrete correlation functions

\begin{equation}
UDCF_{ij}=\frac { (a_i-\bar{a})(b_j-\bar{b}) } { \sigma_a\sigma_b },
\label{UDCF}
\end{equation}

where $\bar{a}$ and $\sigma_a$ are the means and standard deviations of the points $a_i$ in each bin, and in similar manner for series $\mathbi{b}$. Originally, \cite{Edelson1988} used the entire series to calculate the means and the standard deviations, however, this is only valid for the case of stationary time series (mean and standard deviation do not change with time); since AGN light curves rarely behave like that, we calculate the means and standard deviations for the points in individual bins.

\cite{Edelson1988} also suggested that, in order to preserve the proper normalization, it is necessary to replace the $\sigma_a\sigma_b$ in the Eq.~\ref{UDCF} with $[(\sigma_a^2-e_a^2)(\sigma_b^2-e_b^2)]^{1/2}$. \cite{White1994}, however, argue that doing this greatly complicates any direct comparison with the interpolation results, which are unweighted, and therefore this makes an interpretation of the DCF amplitude less straightforward. 

Next, if we have $M$ pairs for which $\tau-\delta t/2 \le \Delta t_{ij} < \tau+\delta t/2$, then the discrete cross-correlation function of $\tau$ is

\begin{equation}
DCCF(\tau)=\frac{1}{M}\sum_{\tau-\delta t/2}^{\tau+\delta t/2}UDCF_{ij}(\Delta t_{ij})
\end{equation}

(Note that the $DCCF(\tau)$ is not defined for a bin with no points).

Unlike the interpolation method, for which the errors in the correlation coefficient are difficult to define, we can define almost directly a standard error for the DCCF. If each of the individual UDCF$_{ij}$ within a single bin were totally uncorrelated, then the standard error in the determination of their mean would be

\begin{equation}
\sigma_{DCCF}(\tau)=\frac{1}{\sqrt{M(M-1)}}\biggl\{\sum_{Bin}[UDCF_{ij}-DCCF(\tau)]^2\biggr\}^{1/2}.
\end{equation}

\subsection{Z-Transformed Discrete Cross-Correlation Function}

The Z-Transformed Discrete Correlation Function by \cite{Alexander1997} is an alternative method for estimating the CCF of sparse, unevenly sampled light curves. The ZDCF corrects several biases of the discrete correlation function method of \cite{Edelson1988} by using equal population binning and Fisher's z-transform \citep[e.g.][and references therein]{Kendall1969,Kendall1973}.

If we have $n$ pairs in a given time-lag bin, the CCF($\tau$) is estimated by the correlation coefficient

\begin{equation}
r=\frac{\sum_{i}^{n}(a_i-\bar{a})(b_i-\bar{b})/(n-1)}{s_as_b},
\end{equation}

where $\bar{a}$, $\bar{b}$ are the bin averages, and $s_a$, $s_b$ are the standard deviations, defined as

\begin{equation}
s_a^2=\frac{1}{n-1}\sum_{i}^{n}{(a_i-\bar{a})^2}.
\label{stddev}
\end{equation}

The sampling distribution of $r$ is highly skewed and far from normal, therefore estimating its sampling error by the simple variance $s_r$ can be very inaccurate.

If $\mathbi{a}$ and $\mathbi{b}$ are drawn from bivariate normal distributions, it is possible to transform $r$ into an approximately normally distributed random variable, Fisher's $z$. Defining

\begin{equation}
z=\frac{1}{2}\ln{\Bigg(\frac{1+r}{1-r}\Bigg)}, \; \; \; \rho=\tanh{z}  ,\ \text{and} \; \; \; \zeta=\frac{1}{2}\ln{\Bigg(\frac{1+\rho}{1-\rho}\Bigg)};
\end{equation}

the mean of $z$ is

\begin{equation}
\bar{z}=\zeta+\frac{\rho}{2(n-1)}\times
\Bigg[ 1+\frac{5+\rho^2}{4(n-1)}+\frac{11+2\rho^2+3\rho^4}{8(n-1)^2} + \cdot\cdot\cdot\Bigg]
\end{equation}

and the variance of $z$ is

\begin{equation}
s_z^2=\frac{1}{n-1}\Bigg[ 1+\frac{4-\rho^2}{2(n-1)}+\frac{22-6\rho^2-3\rho^4}{6(n-1)^2}+\cdot\cdot\cdot \Bigg].
\end{equation}

 Transforming to $r$ again, the interval corresponding to the normal $\pm1\sigma$ error interval can be determined by
 
 \begin{equation}
\delta r_{\pm}= |\tanh{(\bar{z}(r)\pm s_z(r))}-\rho|.
\end{equation}

The binning method for the ZDCF is different from that of \cite{Edelson1988}, in which the binning of the pairs is done in a fixed time interval, i.e. every bin has the same length in time, and the resulting DCCF of that bin is the average of the UDCF of all pairs who fall into the bin; while, the ZDCF binning is for a fixed population, i.e. every bin has the same number of pairs. Each bin contains at least $n_{min}=11$ pairs, which is the minimum number for a meaningful statistical interpretation \citep{Alexander1997}.

Also, in each bin, the interdependent pairs are discarded. This means that in a bin we cannot have two pairs which use the same $a_i$ or $b_j$ element. 
Therefore, light curves with less than 12 points cannot be analyzed by this method.

The lag corresponding to a certain bin is the average of the $\Delta t$ associated with all the pairs in a bin, and in the original method of \cite{Alexander1997}, he proposed to adopt as an uncertainty bar the interval comprising 34\% of the points left and right from the mean. This, however, is not a confidence interval, it is just the interval where 68\% percent of the points used to compute the average lies. However, since the distribution of $\Delta t$ in each bin is not necessarily Gaussian, it can happen that there are not enough points at higher or lower $\Delta t$'s from the mean. Even in some bins, where the $\Delta t$'s were a little separated, the points under or above the average could not complete the 34\%. For these reasons, we decided to use as uncertainty, the entire range covered by the points in the bin, given that we are interested in this range, but it is worth noting that this is not a confidence interval either.

In the original method of \cite{Alexander1997}, the population of the bins begins on $\Delta t = 0$, and then starts adding pairs to a bin going to positive $\Delta t$'s, and in the same way for the negative $\Delta t$'s. However, since the interdependent pairs for each bin are being discarded, this always causes the bins around $\Delta t = 0$, to be really large, which takes away the possibility to obtain (and study) any correlations with a delay close to zero. Therefore, we changed the binning method to start populating bins, starting with the lowest $\Delta t$ of any of the pairs.

\section{Uncertainty in the Delay}

The estimation of the uncertainty in the delay found using cross-correlation, is inconsistent through the literature \citep[e.g.][among others]{Gaskell1987, White1994, Alexander1997}.

For instance, the uncertainty in the ICCF is usually calculated by fitting a gaussian to the peak of the cross-correlation function, and taking the standard deviation as uncertainty; however, this is not a confidence interval (since the cross-correlation function is not a distribution of values); it also does not take into consideration the properties of the individual light curves, which is a parameter that should be taken into account because the sampling of the observed light curve intrinsically limits our temporal resolution. The DCCF uncertainty usually presented in the literature is based on the bin size selected prior to the analysis, which has the same problem as the method used for the ICCF. The error defined for the ZDCF by \cite{Alexander1997} is a fiducial interval; however, a fiducial interval is not the same as a confidence interval. 

We calculated these errors, using Monte Carlo simulations, taking as a starting point the observed light curves used in this work. 
First, we created simulated light curves with the same statistical properties as the ones of the observed light curves (using a random seed), however, with very high time-resolution. This methodology is described in \cite{Timmer1995}, \cite{Schreiber1996}, and in \cite{Emmanoulopoulos2013}. Then, we duplicated one of the simulated light curves and applied a time delay to it. For this work, we tested delays between -250 and 250 days in intervals of 1 day. We re-sampled these two light curves, the original and the delayed, to randomly spaced cadences similar to the one of the observed light curve. This using a random uniform distribution with both light curves being re-sampled with different values. Then, a cross-correlation analysis is performed using the ICCF. The lag found at the maximum of the cross-correlation function, the artificially introduced lag, and the sampling mean for each of the light curves are recorded. This process was repeated until we obtained results from 1.2+ million simulated pairs of light curves. After this, we found the differences between the introduced lags and the obtained lags; these differences were normalized to the sampling means of the light curves. We obtained the distribution of these normalized differences and estimated the 1$\sigma$. These results are presented in Table~\ref{sim_table}.

\begin{table}[htp]
\caption{Results from the simulations to obtain uncertainties on the cross-correlation delays.}
\label{sim_table}
\begin{center}
\begin{tabular}{|c|c|c|}
\hline
Normalization reference & Lower Limit & Upper Limit \\
\hline
Larger Sampling Mean & -0.116 & 0.116 \\
\hline
Smaller Sampling Mean & -0.131 & 0.136 \\
\hline
Interpolation Interval & -0.5 & 0.5 \\
\hline
\end{tabular}
\end{center}
\label{default}
\end{table}%

Since the larger sampling mean and the interpolation interval are symmetrical, and both can be factors to the uncertainty in the lag, we decided to be conservative and obtain the error in the lag as

\begin{equation}
\sigma_{lag} = \sqrt{ (0.5 \times int)^2 + (0.116 \times sm)^2 },
\end{equation}

where $int$ represents the interpolation interval and $sm$ represents the larger sampling mean between the two light curves involved in the cross-correlation.

\section{Cross-Correlation Significance}

For each light curve pair that is being cross-correlated, we generate 50,000 simulated light curves, with the same kernel (statistical properties in the phase space) as the real light curves. Then, we proceed to cross-correlate the 50,000 pairs of simulated light curves. From the resultant cross-correlation functions, we obtain for each lag, the 90, 95 and 99 \% quantiles.

\begin{figure}[!htb]
\begin{center}
\includegraphics[width=0.5\textwidth]{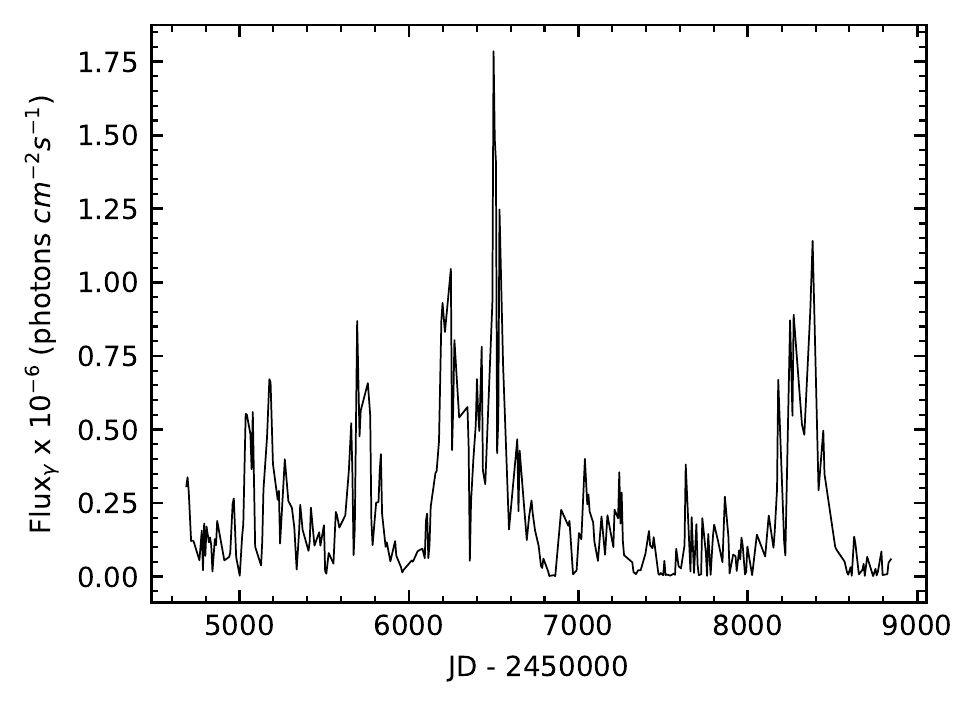}
\includegraphics[width=0.48\textwidth]{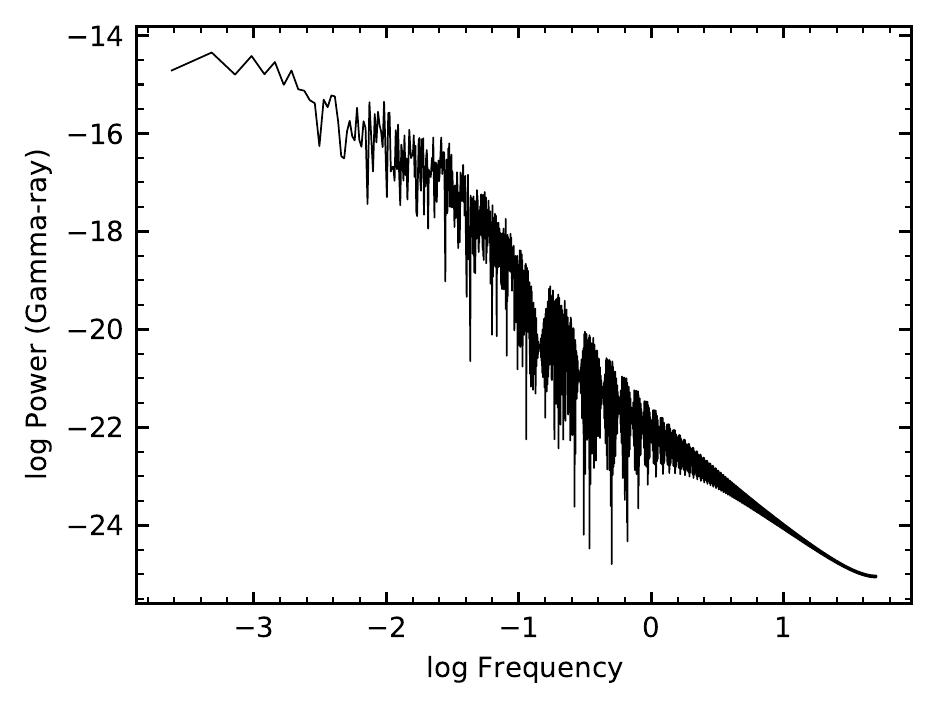}
\caption{Left panel: Gamma-rays light curve of B2 1633+382. Right Panel: Power Spectrum of the light curve shown in the left panel.}
\label{tools_fig1}
\end{center}
\end{figure}

When computing the power spectrum of a light curve, the expected morphology is a power law. For the case of the light curves of B2 1633+382, this is true. An example is shown in Fig.~\ref{tools_fig1}. In order to generate simulated light curves with the same kernel than the observed one, an spectral index is obtained from the power spectrum of each light curve to be analyzed. In this case, the power law spectrum can be written as

\begin{equation}
S(\omega)\sim (1/\omega)^{\beta}.
\label{eq:B1}
\end{equation}

For each frequency, $\omega_i$, we draw two Gaussian distributed random numbers, then multiply them by $\sqrt{\frac{1}{2}S(\omega_i)}\sim (1/\omega)^{\beta /2}$ and use the result as the real and imaginary part of the Fourier transform of the desired data. In the case of an even number of data points, for reasons of symmetry, $f(\omega_{Nyquist})$ is always real. Then only one Gaussian distributed random number has to be drawn.

To obtain a real-valued time series, we choose the Fourier components for the negative frequencies according to $f(-\omega_i)=f*(\omega_i)$, where the asterisk denotes complex conjugation. Finally, we obtain the time series by backward Fourier transformation of $f(\omega)$ from the frequency domain to the time domain.

The generation of the simulated light curves, including the introduction of red noise into the power-law in the phase space, is a combination of the methodologies described in \cite{Timmer1995} and \cite{Schreiber1996}, and is further described in detail in \cite{Emmanoulopoulos2013}.

\bibliography{refs}{}
\bibliographystyle{aasjournal}

\end{document}

%% file: Table1_flux.tex
\begin{table}[htbp]
\centering
\caption{\tablenotemark{*}Sample of the flux measurements for the $\lambda$1350 \AA\ continuum and the C IV $\lambda$1549 \AA\ emission line.}
\begin{tabular}{ccccc}
\hline
\hline
\multirow{2}{*}{JD-2450000} & {Flux Continuum $\lambda$1350 \AA\ } &  {Flux C IV $\lambda$1549 \AA\ } \\
 & $\times10^{-15}\, erg\, s^{-1}cm^{-2}\, {\rm \AA}^{-1}$ & $\times10^{-15}\, erg\, s^{-1}cm^{-2}$ \\
\hline
4745.63 &    12.69 $\pm$    1.62 &    293.74 $\pm$    21.51 \\
4746.63 &    13.31 $\pm$    2.02 &    304.57 $\pm$    26.74 \\
4747.63 &    13.94 $\pm$    2.51 &    255.50 $\pm$    32.97 \\
4748.64 &    15.40 $\pm$    2.71 &    287.53 $\pm$    35.53 \\
4767.60 &    13.64 $\pm$    1.72 &    224.82 $\pm$    22.29 \\
4768.59 &    11.08 $\pm$    0.78 &    233.48 $\pm$    10.09 \\
4769.58 &    12.44 $\pm$    1.38 &    265.99 $\pm$    18.08 \\
4770.58 &    12.26 $\pm$    1.68 &    241.12 $\pm$    22.10 \\
4771.58 &    13.43 $\pm$    1.84 &    269.56 $\pm$    24.21 \\
4772.58 &     9.56 $\pm$    0.82 &    210.70 $\pm$    10.72 \\
\hline
\end{tabular}
\tablenotetext{*}{The complete table is available in a machine-readable form in the online journal.}
\label{Table_flux}
\end{table}

%% file: Table2_cc.tex
\begin{table}
\centering
\caption{Cross-correlation results for the full light curves given in time delays (days) with their uncertainty at 90\% confidence level. All delays have correlations at the $\geq$ 99\% significance level. The results are only shown for the cross-correlation analyses in which the delays are not aliases. All cross-correlations are performed in the order stated in this table e.g. if the delay is positive then the first band leads the second and if the delay is negative the first band lags the second.}

\begin{tabular}{lc}
\hline
\hline
                Bands &         Delay                                 \\
\hline
        1 mm versus X-rays &           -120.4 $\pm$  16.9             \\
        15 GHz versus 1 mm &           -38.9 $\pm$              16.3  \\
     15 GHz versus 1350\AA &  No correlation                          \\
        15 GHz versus C IV &            83.4 $^{+23.9}_{-21.4}$       \\
  15 GHz versus Gamma-rays &           -69.5 $\pm$               8.7  \\
      15 GHz versus X-rays &  No correlation                          \\
1350\AA\ versus Gamma-rays &            -7.4 $\pm$               8.7  \\
      1350\AA\ versus C IV &  No correlation                          \\
    1350\AA\ versus X-rays &  No correlation                          \\
      1350\AA\ versus 1 mm &  No correlation                          \\
    C IV versus Gamma-rays &  No correlation                          \\
        C IV versus X-rays &  No correlation                          \\
          C IV versus 1 mm &  No correlation                          \\
             V versus 1 mm &            26.6 $\pm$              16.3  \\
           V versus 15 GHz &            82.4 $\pm$               4.7  \\
         V versus 1350\AA\ &             0.0 $\pm$               5.9  \\
             V versus C IV &  No correlation                          \\
       V versus Gamma-rays &             0.7 $\pm$               8.7  \\
           V versus X-rays &  No correlation                          \\
    Gamma-rays versus 1 mm &            19.3 $\pm$              16.3  \\
  Gamma-rays versus X-rays &  No correlation                          \\
\hline
\end{tabular}

\label{Table_CC}

\end{table}

%% file: Table3_Fvar.tex
\begin{table}
\centering
\caption{Fractional variability (F$_{\text{var}}$) of the multiple wave bands.}

\begin{tabular}{lc}
\hline
\hline
                Bands &         F$_{\text{var}}$  \\
\hline
         15~GHz &      0.220 $\pm$ 0.001  \\
         1~mm &        0.449 $\pm$ 0.005  \\
         V-band &      0.580 $\pm$ 0.001  \\
        1350\AA\ &     0.486 $\pm$ 0.005 \\
          X-rays &     0.533 $\pm$ 0.008 \\
      Gamma-rays &     1.142 $\pm$ 0.013 \\
\hline
\end{tabular}

\label{Table_Fvar}

\end{table}

%% file: Table4_corr-periods.tex
\begin{table}
\centering
\caption{Linear fitting and Spearman correlation rank test (SCRT) results for the luminosity relations of the full period of observations and of the separated periods defined in Section \ref{sec:var}. The column Continuum refers to the dominant source of UV-continuum in the corresponding period, the flag D refers to disk dominated while J refers to jet dominated. If there is no flag, it means that there is not clear dominant source in the corresponding period. For more details see Section \ref{sec:LumRel}.}
\begin{tabular}{lccccc}
\hline
\hline
                         &       \multicolumn2c{Linear Fit} &        \multicolumn2c{SCRT} & \\
                 Periods &           Slope &        P-value &     $\rho$ &       P-value  & Continuum\\
\hline

  	            Full &     $0.06\pm0.01$ &          1.000 &      -0.10 &         0.045  &  \\

                    Full &     $0.40\pm0.02$ &          1.000 &       0.29 &         0.000  & D\\
                      P1 &     $0.55\pm0.07$ &          0.997 &       0.20 &         0.084  & D\\
		      P3 &     $0.31\pm0.03$ &          1.000 &       0.12 &         0.511  & D\\
		      P5 &     $0.48\pm0.07$ &          0.999 &       0.37 &         0.014  & D\\

                    Full &     $0.14\pm0.03$ &          1.000 &       0.13 &         0.122  & J\\
		      P2 &    $-0.10\pm0.03$ &          0.995 &      -0.47 &         0.002  & J\\
		      P4 &     $0.05\pm0.05$ &          0.221 &       0.13 &         0.350  & J\\
                      P7 &    $-0.05\pm0.09$ &          0.866 &      -0.22 &         0.253  & J\\

		      P6 &    $-0.10\pm0.03$ &          1.000 &      -0.22 &         0.037  &  \\
		      
\hline
\end{tabular}
\label{Table_corr-periods}
\end{table}